\documentclass[%
 reprint,
 amsmath,amssymb,
 aps,
]{revtex4-2}

\usepackage{graphicx}
\usepackage{dcolumn}
\usepackage{bm}
\usepackage{comment}

\begin{document}

\preprint{APS/123-QED}

\title{Causal Spike Timing Dependent Plasticity Prevents \\ Assembly Fusion in Recurrent Networks}

\author{Xinruo Yang}
\affiliation{
    Department of Neurobiology\\
    Grossman Center for Quantitative Biology and Human Behavior \\
    University of Chicago, Chicago, IL 60637, USA\\ 
}
\email{xinruoyang@uchicago.edu} 

\author{Brent Doiron}
\affiliation{
   Departments of Neurobiology and Statistics\\
    Grossman Center for Quantitative Biology and Human Behavior \\
    University of Chicago, Chicago, IL 60637, USA\\ 
}
\email{bdoiron@uchicago.edu}


\date{\today}

\begin{abstract}
The organization of neurons into functionally related assemblies is a fundamental feature of cortical networks, yet our understanding of how these assemblies maintain distinct identities while sharing members remains limited. Here we analyze how spike-timing-dependent plasticity (STDP) shapes the formation and stability of overlapping neuronal assemblies in recurrently coupled networks of spiking neuron models. Using numerical simulations and an associated mean-field theory, we demonstrate that the temporal structure of the STDP rule, specifically its degree of causality, critically determines whether assemblies that share neurons maintain segregation or merge together after training is completed. We find that causal STDP rules, where potentiation/depression occurs strictly when presynaptic spikes precede/proceed postsynaptic spikes, allow assemblies to remain distinct even with substantial overlap in membership. This stability arises because causal STDP effectively cancels the symmetric correlations introduced by common inputs from shared neurons. In contrast, acausal STDP rules lead to assembly fusion when overlap exceeds a critical threshold, due to unchecked growth of common input correlations. 
Our results provide theoretical insight into how spike-timing-dependent learning rules can support distributed representation where individual neurons participate in multiple assemblies while maintaining functional specificity.

\end{abstract}

\maketitle


\section{Introduction}

The synaptic connections between neurons are not arbitrarily distributed, and rather they reflect an organization that supports brain network function. The pioneering work of Donald Hebb~\cite{hebb1949organization} proposed that a population of neurons that repeatedly or persistently fire together will form a strongly interconnected group, often termed a neuronal assembly \cite{harris2013cortical,markram2011history}. The strong recurrent connectivity within an assembly acts to reinforce neuronal response, buffering against the inherent unreliable nature of single neuronal activity \cite{tang2018recurrent}. This is one reason why neuronal assemblies serve as essential building blocks of many theories of neural computation \cite{buzsaki2010neural,harris2013cortical,papadimitriou2020brain} and associative memory \cite{neves2008synaptic,miehl2023formation}. 
Over the past two decades significant experimental evidence has surfaced to support the existence of assembly structure in cortical networks \cite{yoshimura2005excitatory,perin2011synaptic,holtmaat2016functional,zaki2024offline}. In particular, in rodent visual cortex there is clear evidence of stronger and more probable connections between neurons with similar orientation tuning \cite{ko2011functional,cossell2015functional,lee2016anatomy}, indicating that assembly structures and neuron functions are related. A longstanding goal in neuroscience is to identify the synaptic learning mechanisms that underlie assembly formation. 

Many early modeling studies focused on firing rate based plasticity rules for synaptic wiring \cite{bienenstock1982theory,magee2020synaptic,graupner2016natural}. Later experimental studies revealed that the spike time correlations between pre- and post-synaptic neurons over fine timescales (10s of milliseconds) also play an important role in synaptic learning \cite{bell1997synaptic, markram1997regulation, markram2012spike, bi1998synaptic,caporale2008spike}. These recordings of spike timing dependent plasticity (STDP) uncovered a temporally causal rule for synaptic learning (often termed Hebbian), where if a pre-synaptic neuron spikes before a post-synaptic neuron spikes then the synapses is potentiated (strengthened), otherwise the synapse depressed (weakened). Theoretical work first explored a natural consequence of this causal rule and established that feedforward chains of synaptically wired neurons would easily develop \cite{masuda2007formation,takahashi2009self,ravid2016shaping,fiete2010spike}. However, a feedforward chain is an architecture that is quite distinct from a recurrently coupled neuronal assembly.

Later theoretical work considered recurrently coupled networks with causal STDP learning. When pairs of neurons are considered then if the STDP rule is potentiation dominated then strong recipricol connections can develop \cite{babadi2013pairwise}. In large networks, strong recurrently coupled assemblies can be trained even with depression dominated STDP, due to synchronous common inputs to pairs of neurons within an assembly driving temporally selective synaptic potentiation \cite{ocker2019training}. Associated mean field theory shows how assembly structure can be self-reinforcing, so that trained synaptic structures are stable post training \cite{ocker2015self,ocker2019training,manz2023purely}. These studies show how temporally precise STDP can be used to not only embed feedforward wiring, but also can support the formation of recurrently coupled neuronal assemblies.

However, many of these studies only consider assembly learning with no overlap in membership -- meaning that a neuron is a part of only one assembly. This assumption is at odds with the clear evidence that individual neurons participate in the representation of multiple distinct stimuli, and exhibit a mixed or varied selectivity in their responses \cite{rigotti2013importance, cai2016shared, wenzel2021identification,gastaldi2021shared}. Thus, if each stimulus has an associated assembly and neurons are tuned for multiple stimuli it is expected that assemblies will share members. 
This overlap is essential for networks to be able to store a large number of distinct stimuli in its recurrent synaptic coupling \cite{aljadeff2021synapse,fusi2021memory}. Understanding how STDP affects stable assembly formation when there is significant overlap is a very understudied problem and there remains much to explore.  

The temporal selectivity of STDP is quite diverse, with many biological components controlling how pre- and post-synaptic spike timing translate into potentiation and depression \cite{brzosko2019neuromodulation,feldman2012spike}. Indeed, there is a significant range of STDP learning rules, with some being of a classic causal form, while others show a symmetric rule where the ordering of pre- and post-synaptic spike times is unimportant and only their temporal coincidence matters \cite{graupner2012calcium}. Historically, causal STDP have been the focus of many theoretical studies \cite{ocker2015self,ocker2019training,masuda2007formation,kempter1999hebbian,babadi2013pairwise,ravid2016shaping}. However, recently Manz and colleagues \cite{manz2023purely} studied the formation of stable assembles with membership overlap in networks with acausal (symmetric) excitatory STDP. They showed how assemblies could remain segregated with limited assembly overlap.  Our study explores how the degree of causality in STDP affects the tolerance for assembly overlap.  Through the derivation of self-consistent mean field equations for weight dynamics, we show how networks with causal/asymmetric STDP are far more tolerant of assembly overlap than networks with acausal/symmetric STDP. In this way, our work then extends the functionality of causal STDP to beyond simply allowing feedforward structure to develop, yet to also greatly enhancing the assembly capacity in recurrent networks.

\section{Results}

\subsection{Model Framework}

\subsubsection{Neuronal Dynamics}

We consider a network of $N$ spiking neuron models, with $N_E=4N/5$ of the neurons being excitatory ($E$) and the remaining $N_I=N/5$ of them being inhibitory ($I$). Let $\gamma_E=N_E/N=4/5$ and $\gamma_I=N_I/N=1/5$ denote the fractions of excitatory and inhibitory neurons, respectively.  The membrane potential of neuron $i$ in population $\alpha \in \{E,I\}$, denoted as $V^i_\alpha$, obeys an exponential integrate-and-fire model formalism \cite{fourcaud2003spike}:
\begin{widetext}
\begin{equation}
    \tau_{\alpha} \frac{dV^i_\alpha}{dt} = (E_L-V^i_\alpha)+\Delta_T \exp\Big(\frac{V^i_\alpha-V_T}{\Delta_T}\Big)+ I^i_\alpha(t)+\sum_{\beta=E,I}\sum_{j=1}^{N^\beta}W_{\alpha\beta}^{ij}(t)\cdot \big(J_{\textrm{syn}} * y^j_\beta(t)\big) . \label{eq:ch2_EIF}
\end{equation}
\end{widetext}
The passive properties of the neuron are given by the leak reversal potential $E_L$ and membrane time constant $\tau_{\alpha}$. Action potential initiation is modeled by the exponential term in Eq.~\eqref{eq:ch2_EIF} where parameter $V_T$ sets the membrane value where the action potential upstroke is approximately triggered, and $\Delta_T$ is a scaling parameter.
Spike and reset dynamics are given by the discontinuity $V^i_\alpha (t_\alpha^{ik})=V_s \to V^i_\alpha ([t_\alpha^{ik}]_+) = V_r < V_s$, where $t^{ik}_\alpha$ is the $k^{\textrm{th}}$ time neuron $i$ from population $\alpha$ crosses the spike threshold $V_s$ and is subsequently reset to $V_r$ and held there for a refractory period $\tau_{\textrm{ref}}$ (Fig. \ref{eq:ch2_EIF}C). The spike train for neuron $i$ in population $\alpha$ is given by $y^i_\alpha(t)=\sum_k \delta(t-t^{ik}_\alpha)$. The third term in Eq.~\eqref{eq:ch2_EIF}, $I^i_\alpha (t)$, denotes a background external input to neuron $i$, which is modeled by an associated diffusion approximation \cite{brunel2000dynamics, gerstner2014neuronal, gluss1967model, johannesma1968diffusion, capocelli1971diffusion}:
\begin{equation}  \label{eq:ch2_Iext}
    I^i_{\alpha}(t) = \mu_{\alpha}^{\textrm{ext}} + \sigma_{\alpha}^{\textrm{ext}}\xi_{\alpha}^i(t).
\end{equation} 
Here $\xi_{\alpha}^i(t)$ is a Gaussian white noise stochastic process characterized by expectation $\langle \xi^i_{\alpha}(t)\rangle=0$ and correlation function $\langle \xi^i_\alpha(t)\xi^j_\beta(t')\rangle=\delta_{\alpha\beta}\delta^{ij}\delta(t-t')$, where $\delta^{ij}$ and $\delta_{\alpha\beta}$ are Kronecker functions ($\delta_{\alpha\beta}=1 \leftrightarrow \alpha=\beta$; $\delta^{ij}=1 \leftrightarrow i=j$) and $\delta(t-t')$ is a Dirac function. Finally, the fourth term in Eq.~\eqref{eq:ch2_EIF} is the recurrent input from the other neurons in the network, where $J_{\rm{syn}}(t)=\frac{1}{\tau_{\rm{syn}}}\exp(-t/\tau_{\rm{syn}})H(t)$ is an exponential synaptic filter describing the dynamics of the post-synaptic current, with $*$ denoting temporal convolution and $H(t)$ being the Heaviside function ($H(t)=1$ for $t>0$ and $H(t)=0$ for $t\le 0$). The coefficient $W^{ij}_{\alpha\beta}$ is the synaptic strength from neuron $j$ in population $\beta$ to neuron $i$ in population $\alpha$.

All neuronal parameters used in our study are given in Table \ref{tab:neuronal} (see Methods section \ref{parameters}).

\subsubsection{Excitatory Synaptic Weight Dynamics}

In this network we consider the well-studied phenomenological model of spike time dependent plasticity (STDP), where the synapse strength $W^{ij}_{\alpha\beta}(t)$ changes depending on the difference between pre- and post-synaptic spike times \cite{ocker2015self, ocker2019training, akil2021balanced,gerstner2014neuronal,abbott2000synaptic}. We will only consider the $E \to E$ and the $I \to E$ synapses as plastic.

For each pair of excitatory pre- and post-synaptic spike times $(t_{ \textrm{pre}}=t^{jk},t_{\rm{post}}=t^{il})$ with time lag $s=t_{\rm{post}}-t_{\rm{pre}}$, where $t^{jk}$ indicates the time of $k$-th spike of neuron $j$, the synapse evolves as:
\begin{equation*}
W^{ij}_{EE}\longrightarrow W^{ij}_{EE}+L(s).
\end{equation*}
 Here $L(s)$ is the STDP learning rule. We omit the subscript $\alpha\beta$ notation in $W^{ij}$ for the convenience of presentation.

We focus on two distinct $E \to E$ learning rules. The first (Fig.~\ref{fig:Fig1}C) is a temporally causal $E\to E$ rule (sometimes referred to as Hebbian), $L^{\textrm{c}}(s)$, which is defined as follows: 
\begin{equation*}
  L^{\textrm{c}}(s)=\begin{cases}
    f^c_{+} \exp\Big({-\frac{s}{\tau_{+}^c}}\Big), & \text{if $s>0$}.\\
    -f_{-}^c \exp\Big({\frac{s}{\tau_{-}^c}}\Big), & \text{if $s\leq 0$}.
  \end{cases}
\end{equation*}
The parameters $\tau_+^c>0$ and $\tau_-^c>0$ represent the time scales of the forward and backward decay of the causal STDP rule, while $f_{+}^c>0$ and $f_{-}^c>0$ denote the maximal amplitudes of each synaptic change, respectively. This rule is termed causal because pre-synaptic spikes that precede post-synaptic spikes ($s>0$) result in synaptic potentiation ($L^c(s)>0$), whereas if they follow post-synaptic spikes ($s<0$), synaptic depression occurs ($L^c(s)<0$).  

The second learning rule, $L^{ac}(s)$, is temporally acausal and obeys: 
\begin{equation*}
  L^{ac}(s)=\begin{cases}
    f_{+}^{ac} \exp\Big({-\frac{s}{\tau_+^{ac}}}\Big)\cdot (T_+-s), & \text{if $s>0$}\\
    f_{-}^{ac} \exp\Big({\frac{s}{\tau_-^{ac}}}\Big)\cdot (T_- + s), & \text{if $s\leq 0$}.
  \end{cases}
\end{equation*}
Here the parameters $f_+^{ac}>0$, $f_-^{ac}>0$, $\tau_+^{ac}>0$ and $\tau_-^{ac}>0$ serve the same roles as in $L^c(s)$, and the $(T_{\pm} \mp s)$ terms biases the plasticity to depression for large $\vert s \vert $.  
The learning rule is acausal because the relative timing of the pre- and post-synaptic spike times is less important, and synaptic potentiation occurs for small $\vert s \vert $ and depression for larger  $\vert s \vert $.  

In this study, we operate in a weak coupling regime where $W_{\alpha\beta}^{ij}$ scales as $ 1/N\equiv \epsilon$. 
We assume that the integral of the STDP function $L^{\textrm{c}}(s)$ is small compared to its $L_1$ norm: 
\begin{equation*}
    \int L^c(s)ds \sim \mathcal{O}(\epsilon) \cdot ||L^c(s)||_1.
\end{equation*} \label{eq:ch2_balance_STDP}
This means that the depression and potentiation components of $L^c(s)$ roughly cancel each other out when $L^c(s)$ and $L^{ac}(s)$ is integrated. This assumption ensures that the plasticity will not be dominated by chance spike coincidences of pre- and post-synaptic neurons \cite{ocker2019training, miehl2023formation}.

\subsubsection{Inhibitory Synaptic Weight Dynamics}

Additionally, we assume the connections from $I\to E$ connections are plastic. Their plasticity is described by the homeostatic learning rule $L^h(s)$ \cite{vogels2011inhibitory, froemke2007synaptic}:
\begin{equation}\label{eq:L_I}
    \begin{split}
        L^h(s) & = f^h \exp\Big({-\frac{|s|}{\tau^h}}\Big),\\
    \end{split}
\end{equation}
with $f^h<0$ represents the maximal amplitude change of the inhibitory synapse corresponding to pre- and post-synaptic spikes with time lag $s$. In addition to the synaptic update corresponding to spike pairs, each pre-synaptic (inhibitory) spike drives depression of the inhibitory synapses by $d^h>0$ \cite{woodin2003coincident, kilman2002activity}:
\begin{equation*}
    W^{ij}_{EI} \rightarrow W^{ij}_{EI} + d^h.
\end{equation*}
We note that for $I \to E$ connections we have $W^{ij}_{EI} <0$ so that adding $d^h>0$ leads to a weakening of the synapse. 
Previous studies \cite{vogels2011inhibitory, ocker2019training} have shown that this term gives the inhibitory synapse a drift such that the excitatory neurons would eventually fire at the target firing rate, $r_E^{\textrm{target}} = -d^h/(2 f^h \tau^h)$. $r_E^{\textrm{target}}>0$ is ensured by $d^h>0$ and $f^h<0$. This rule was found to stabilize the activity of the target excitatory neurons by balancing their excitatory and inhibitory input \cite{vogels2011inhibitory, ocker2019training}. This mechanism is further explained in the Methods sec.~\ref{appendix:inhibitory_plasticity}.

All plasticity parameters used in our study are given in Table \ref{tab:plasticity} (see Methods section \ref{parameters}).

\subsection{The Fate of Assembly Structure }

\setlength\belowcaptionskip{-3ex}
\begin{figure}[t!]
    \begin{center}
    \includegraphics[scale=1]{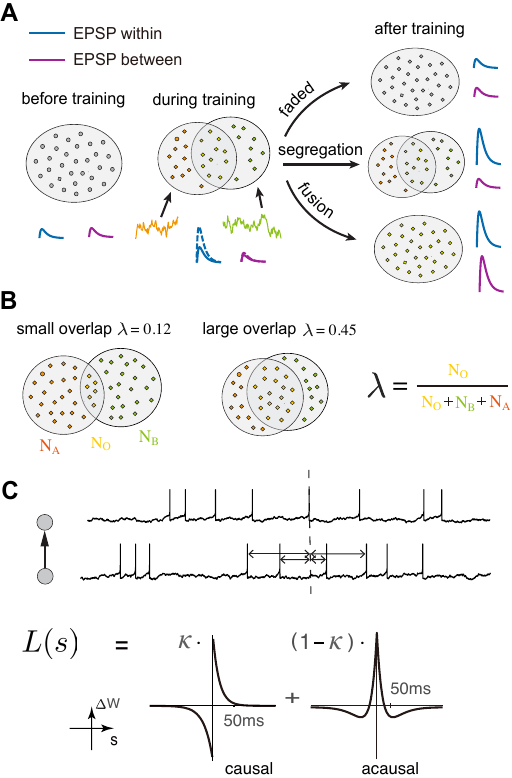} 
    \end{center}
    \vspace{-4mm}
    \caption{(A) Illustration of three possible assembly fates after training. We create two assemblies with overlap using two uncorrelated signals, and there are three possible dynamics after training. 1) The assembly structure fades away and all synaptic connections depress to their original pre-training values. 2) The assemblies remain segregated with strong within assembly connections and weak between assembly connections. 3) The assemblies fuse into one large assembly with strong connections throughout. (B) Definition of the overlap ratio $\lambda$ (C) Illustration of the $\rm{E} \to \rm{E}$ STDP rule with causality parameter $\alpha$.}
\label{fig:Fig1}
\end{figure}

The base network connectivity in our model is unstructured before training. Specifically, the adjacency matrix (existence of a connection) for the connectivity from pre-synaptic population $\beta$ to the post-synaptic population $\alpha$ is generated via the simple rule:
\begin{equation}
    G^{ij}_{\alpha\beta} \sim \textrm{Bernoulli}(p_{\alpha\beta}). \label{eq:A}
\end{equation}
Further, we initialize the synaptic weights to be homogeneous within a connection class:  
\begin{equation}
    W^{ij}_{\alpha\beta}(t=0) = \epsilon \cdot G^{ij}_{\alpha\beta}J_{\alpha\beta}. \label{eq:Wint}
\end{equation}
The combination of Eqs. \eqref{eq:A} and \eqref{eq:Wint} is such that there is no initial assembly structure pre-wired within the network.

To embed assembly structure in the network we will follow \cite{ocker2015self, ocker2019training} and use a common dynamic `training' signal given to subgroups of neurons with the aim of strengthening within group synaptic connectivity.
Throughout our study we consider two distinct training signals, $s_A(t)$ and $s_B(t)$, and assign $E$ neurons to receive either signal $A$, signal $B$, or both signals. We accomplish this by augmenting the external input during training to be:
\begin{equation*}
    I^i_E(t)=\mu_E^{\rm{ext}} + \sigma_E\xi_E^i(t)+M^i_A s_A(t)+M^i_B s_B(t).
\end{equation*}
Here $\mu_E^{\rm{ext}}$ and $\sigma_E\xi^i_E(t)$ set the background state as defined in Eq.~\eqref{eq:ch2_Iext}. The two signals $s_A(t)$ and $s_B(t)$ are also modeled as white noise fluctuations, yet are a common source of fluctuations across all neurons that are members of an intended assembly. 
During training if the $i^{\textrm{th}}$ $E$ neuron is intended to be a member of assembly $A$ (or $B$) then $M^i_A$ (or $M^i_B$) is set to 1.

The central question of our study is: what is the fate of assembly structure after training has finished? Before training, connection strengths are small and not organized into any group structure (Fig.~\ref{fig:Fig1}A, left). During training, the STDP rules will promote strong within group connectivity, leaving between group connectivity weak (Fig.~\ref{fig:Fig1}A, middle). After training has stopped, there are three possible fates for assembly structure (Fig.~\ref{fig:Fig1}A, right).  First, the assemblies may dissolve and all connections will depress to match the strengths before training. Second, the assemblies will remain segregated, with the within assembly connectivity persisting at high levels, while the between assembly connections remain weak. Third, the assemblies may fuse together and all connections, within and between, potentiate to high levels. The segregated case represents stable memory formation as the structure embedded by the training phase persists post training. The faded or fused assemblies represent failed learning, since the end state of the network shows no evidence of the training inputs.     

To investigate assembly fate we will consider two control parameters of our network.  The first is the degree of assembly overlap (Fig.~\ref{fig:Fig1}B). When assigning neurons to assembly membership a fraction of neurons will have both $M^i_A=M^i_B=1$. Let the number of $E$ neurons that are members of both assemblies be $N_O$, and each assembly have $N_X$ neurons that are exclusive members (we assume equally sized assemblies). With each neuron assigned to at least one assembly we have $N_E=2N_X+N_O$. The overlap parameter is then:
\begin{equation*}
    \lambda = \frac{N_O}{N},
\end{equation*}
and $\lambda$ ranges from no overlap ($\lambda=0$) to  complete overlap ($\lambda=\gamma_E=0.8$, where $\gamma_E$ is the fraction of excitatory neurons). 

The second parameter $\kappa$ controls the degree of causality in the STDP learning rule (Fig.~\ref{fig:Fig1}C). Specifically, the $E\to E$ plasticity is given by:  
\begin{equation*}
    L(s)=\kappa L^c(s)+(1-\kappa)L^{ac}(s).
\end{equation*}
With $\kappa=0$ the rule is completely acausal, while for $\kappa=1$ the rule is fully causal. 

It is natural to expect that as the overlap parameter $\lambda$ grows the possibility of assembly fusion increases, since for large $\lambda$ the assemblies will share many members and consequently interact more strongly. Thus, by varying $\lambda$ we will have an expected influence on assembly fate. However, it is less obvious how assembly fusion versus segregation will be affected by the degree of causality $\kappa$ in the STDP rule $L(s)$. Classically, causality in $L(s)$ promotes unidirectional connections \cite{song2000competitive, abbott2000synaptic, song2001cortical}: if neuron $j$ tends to lead neuron $i$ in spiking ($s>0$) then potentiation in $W^{ij}$ occurs, but this necessarily causes any backward connection $W^{ji}$ to depress since $i$ must lag $j$ ($s<0$). This feature of causal STDP is ideal to prevent assembly fusion, since unwanted reciprocal connections between assemblies will be suppressed. However, this same logic is at odds with the formation of strong reciprocal connections within an assembly in the first place. The central goal of our study is to characterize the boundary between assembly segregation and assembly fusion defined in $(\kappa,\lambda)$ parameter space.

\setlength\belowcaptionskip{-3ex}
\begin{figure}
    \begin{center}
    \setlength{\belowdisplayskip}{3pt}
    \setlength\abovedisplayskip{0pt}
    \includegraphics[scale=1.0]{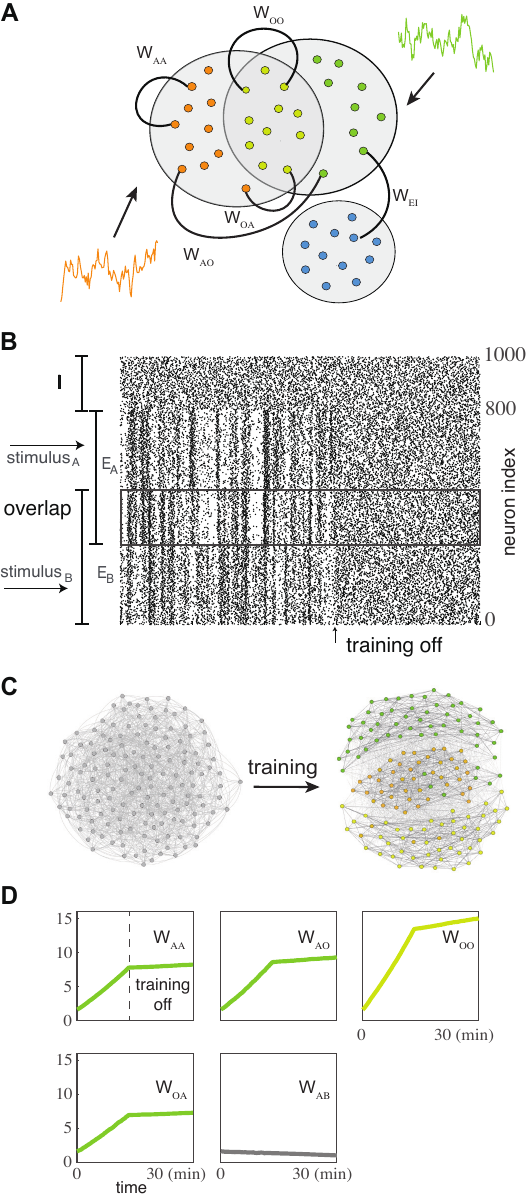}
    \end{center}
    \vspace{-4mm}
    \caption{Illustration of the network model and training protocol (A) We consider an $E - I$ network model with two assemblies ($A$ and $B$) with overlap ($O$) in the excitatory population due to common training signals. The set $\{W_{AA},W_{AB}, W_{AO}, W_{OA}, W_{AI} \}$ of the key plastic weights are marked.  (B) A raster plot of spiketimes of all neurons during and after training (C) Visualization of the connectivity between a subset of the excitatory neurons (D) Dynamics of the mean averaged weights during and after training.} \label{fig:2}
\end{figure}

\subsection{Numerical Exploration of Assembly Fusion or Segregation.}

To begin we divide our network into four sub-populations defined by the training phase (Fig.~\ref{fig:2}A). Neurons that receive only stimulus $A$ or $B$ are identified as population $A$ or $B$, respectively. Neurons that receive both stimuli are identified as the overlap population $O$. Finally, the inhibitory neurons (which do not receive a training signal) form the last population $I$.  We separate $E$ neurons in the overlap from the remaining neurons in the assemblies since their evolution is distinct from that of non-overlap $E$ neurons. After sufficient training (Fig.~\ref{fig:2}B) the $E$ network organizes itself into clearly demarcated $A$, $B$, and $O$ populations (Fig.~\ref{fig:2}C). 

The network dissection into $\{A,B,O,I\}$ prompts us to define normalized mean variables for the dynamic weights in the network:
\begin{equation*}
    W_{\alpha\beta} = \frac{1}{N_{\alpha}N_{\beta}}\sum_{i\in \alpha, j\in \beta} \frac{W^{ij}_{\alpha\beta}}{\epsilon}\sim\mathcal{O}(1),  
\end{equation*}
where $\alpha,\beta\in\{A,B,O,I\}$. Because of the symmetry between the $A$ and $B$ populations only the mean weight vector ${\bf W_m}=[W_{AA}, W_{AB}, W_{AO}, W_{OA}, W_{OO}]$ need be considered (Fig.~\ref{fig:2}B). During training the recurrent synapses within the assemblies show rapid growth (Fig.~\ref{fig:2}D; $W_{AA}$, $W_{OA}$, $W_{AO}$, $W_{OO}$), while the between assembly weights remain small (Fig.~\ref{fig:2}D; $W_{AB}$). The rapid synaptic growth for within assembly coupling is due to the strong correlating force of the training signals. 
After training the spike train correlations become much smaller, and consequently the evolution of the synaptic weights slows significantly. However, the basic structure of a strong within ($W_{AA}$) and weak between ($W_{AB}$) assembly connectivity persists. This is the key signature of stable assembly segregation. 

\noindent 

\begin{figure*}
    \begin{center}
    \includegraphics[scale=1]{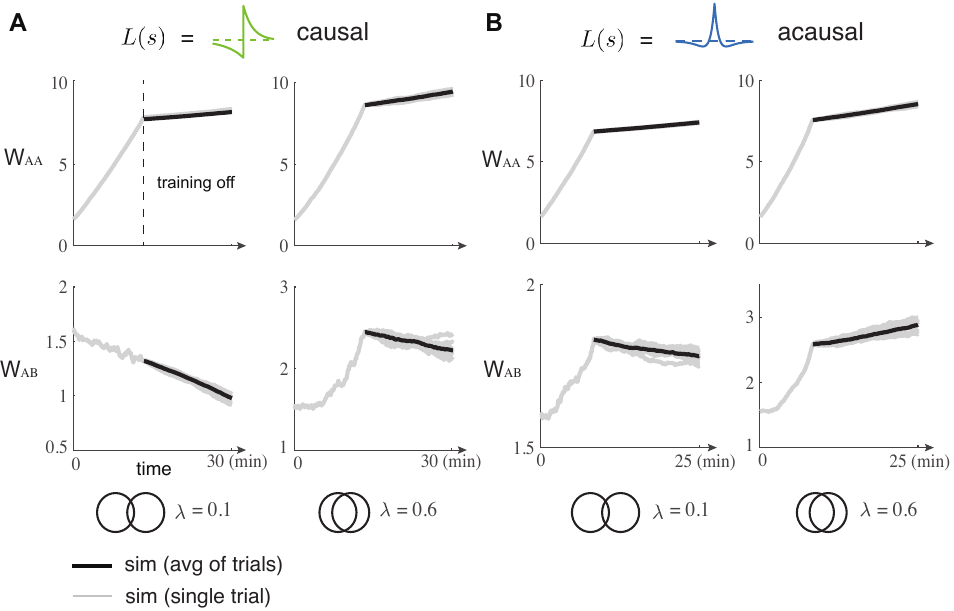}
    \end{center}
    \vspace{-4mm}
    \caption{The dynamics of the mean weights between neuron pairs in the same assembly, $W_{AA}$, and neuron pairs from different assemblies, $W_{AB}$, during and after training for small overlap $\lambda=0.1$ and large overlap $\lambda=0.6$. (A) For the causal STDP rule ($\kappa=1$) , the assemblies remain segregated post training ($W_{AA}$ increases while $W_{AB}$ decreases)  for both small and large overlap. (B) For the acausal STDP rule ($\kappa=0$) the assemblies fuse (both $W_{AA}$ and $W_{AB}$ increase) for large overlap. The grey lines consists of the same realization of training and twenty realizations of after training dynamics, the black line represents the averaged after training dynamics.} \label{fig:ch3_fig3}
\end{figure*}

These initial results prompt us to investigate the $(\kappa,\lambda)$ dependence of the dynamics of $(W_{AA}(t), W_{AB}(t))$ post training.
To begin, we consider the fully causal STDP learning rule ($\kappa=1)$. For both small $(\lambda=0.1)$ and large $(\lambda=0.6)$ assembly overlap the mean weight dynamics post-training preserve the assembly structure embedded during training (Fig.~\ref{fig:ch3_fig3}A). This is clear since for both small and large $\lambda$ the mean within assembly synaptic weight $W_{AA}$ continues to grow after training has ceased, while the mean between assembly weight $W_{AB}$ continues to decay. By contrast, a fully acausal STDP learning rule ($\kappa=0$) and large $\lambda$ lead to assembly fusion (Fig.~\ref{fig:ch3_fig3}B). These numerical results show that the temporal structure of the learning rule $L(s)$ determines the assembly fate post training. To deepen our understanding of this dependence we next develop a mean field theory for the mean synaptic weights ${\bf W_m}$.

\subsection{Mean Field Theory of Synaptic Weight Dynamics} \label{sec:derivation}

 Following past work \cite{ocker2015self,ocker2019training,gilson2010emergence,babadi2013pairwise,kempter1999hebbian}, we separate the slow time scale of weight dynamics from the fast time scale of the cross-covariance of neuron spike trains to derive a set of self consistent differential equations for the slow weight evolution.

For our pair-based learning rule $L(s)$, the weight change $\Delta W^{ij}_{\alpha\beta}$ over an interval $(t, t+T)$ follows \cite{kempter1999hebbian}:
\begin{equation} \label{eq:delta W}
    \Delta W^{ij}_{\alpha\beta} = \int_{t}^{t+T} L(s) \big[ y^i_\alpha(s) y^j_\beta(t+s) \big] ds,
\end{equation}
where $y^i_\alpha(t)$ is the spike train of neuron $i$ in population $\alpha$. If we take the weight update to be small (i.e. $f_+, f_- << \mathcal{O}(\epsilon)$) then $\vert \Delta W^{ij}_{\alpha\beta}\vert  << \vert W^{ij}_{\alpha\beta} \vert $ and the evolution of $W^{ij}_{\alpha\beta}$ occurs on a slow timescale, typically minutes to hours. On this slow timescale the weight dynamics obey \cite{kempter1999hebbian}:
\begin{equation}
    \frac{d W^{ij}_{\alpha\beta}}{dt} = \int_{-\infty}^{\infty} L(s) \big[r^i_\alpha r^j_\beta + C^{ij}_{\alpha\beta}(s)\big]ds. \label{eq:kemp_text}
\end{equation}
Here $ r^i_\alpha=\langle y^i_\alpha(t) \rangle $ is the mean firing rate of neuron $i$ in population $\alpha$, so that $r^i_\alpha r^j_\beta$ captures chance spike correlations between neurons $i$ in population $\alpha$ and $j$ in population $\beta$. The temporal component of learning depends on the cross-correlation function between spike trains $y^i_\alpha (t)$ and $y^j_\beta(t)$ given by $C^{ij}_{\alpha\beta}(s)=\langle (y^i_\alpha(t)-r^i_\alpha)(y^j_\beta(t+s)-r^j_\beta)\rangle $. The function $C^{ij}_{\alpha\beta}(s)$ describes the mean corrected expectation of neuron $j$ spiking at a time interval $s$ after ($s>0$) or before ($s<0$) a spike from neuron $i$. Eq. \eqref{eq:kemp_text} describes how the slow dynamics of weight dynamics depend on the statistics of fast timescale spiking activity.  

On the fast time timescale of neuronal spiking, we consider the recurrent synaptic connectivity ${\bf W}$ to be fixed; ${\bf W}$ is the full $N \times N$ connectivity matrix. To determine spike train correlations from network connectivity, we employ linear response theory in the asynchronous firing regime \cite{lindner2005theory, trousdale2012impact, hu2013motif, ocker2015self, ocker2019training}. In this regime, the cross-correlation is computed as:
\begin{equation*}
    \mathbf{C} = (\mathbf{I}- \mathbf{W} \circ \mathbf{K})^{-1} \mathbf{C}_0 (\mathbf{I}- \mathbf{W}^{*} \circ \mathbf{K}^{*})^{-1},
\end{equation*}
where $\mathbf{C}$ is the cross-correlation matrix with element $C^{ij}(s)$, $\mathbf{C}_0$ is the auto-correlation matrix of neurons in steady state, and $\mathbf{K}$ describes neural response properties. A complete derivation is given in Methods sec.~\ref{diffusion_methods}.

In our parameter regime of weak connections, we always have the eigenspectrum of the effective matrix satisfies $\rho(\mathbf{W} \circ \mathbf{K})<1$, so we can expand the inverse matrix (and its complex conjugate) as its Neumann's Series:
\begin{equation} \label{eq:expansion_text}
    (\mathbf{I}- \mathbf{W} \circ \mathbf{K})^{-1} = \sum_{k=0}^{\infty}(\mathbf{W} \circ \mathbf{K})^k.
\end{equation}
Inserting the expansion Eq.~\eqref{eq:expansion_text} into Eq.~\eqref{eq:kemp_text}, grouping terms by connectivity motifs $q^{\textrm{motif}}_{\alpha\beta}(\mathbf{W_m})$ and taking the mean over synapses yields mean-field equations for population-averaged weights:
\begin{equation} \label{eq:dW/dt full text}
\frac{dW_{\alpha\beta}}{dt} = \sum_{\textrm{motif}} S^{\textrm{motif}}_{\alpha\beta}  q^{\textrm{motif}}_{\alpha\beta}(\mathbf{W_m}).
\end{equation} 
The coefficient $S^{\text{motif}}$ quantifies a motif's influence on plasticity through the integration:
\begin{equation} \label{eq:S_text}
S^{\text{motif}}_{\alpha\beta} = \int_{-\infty}^{\infty} L(s) M^{\text{motif}}_{\alpha\beta}(s)ds,    
\end{equation}
where $M^{\text{motif}}_{\alpha\beta}(s)$ represents the response kernel characterizing how the motif affects the connection $W_{\alpha\beta}$. 

The motif kernels $M^{\text{motif}}_{\alpha\beta}(s)$ depend on  the cellular/synaptic response $K_\alpha$ and the auto-covariance $C^0_{\alpha}$ (see Methods sec.~\ref{diffusion_methods}). The homeostatic rule for inhibitory plasticity forces the excitatory neurons to fire at the target firing rate $r_{E}=r_{E}^{\rm{target}}$ \cite{vogels2011inhibitory, ocker2019training}. This means the response kernels $K_{\alpha}\approx K_{E}$ and the auto-covariance $C^0_{\alpha}\approx C^0_{E}$ are approximately the same for all the excitatory neurons, irrespective of their assembly association. Thus, the motif kernels $M^{\text{motif}}_{\alpha\beta}(s)$ do not depend on $\alpha,\beta \in \{A,B,O\}$ in the excitatory populations, and rather can all be described by $\alpha=\beta=E$. 

An important observation is the linear dependence of the motif coefficients $S^{\rm motif}$, and consequently the weight dynamics $dW_{\alpha\beta}/dt$, on the STDP rule $L(s)$, as evident from Eqs.~\eqref{eq:dW/dt full text} and \eqref{eq:S_text}. By definition, $L(s)$ scales with small weight updates ($f_+,f_-$) over short timescales and with the weights themselves over long timescales. Therefore, when Eq.~\eqref{eq:dW/dt full text} is computed over long timescales, $S^{\rm motif} \sim \mathcal{O}(1/N)$. However, due to the linear dependency, we can introduce a renormalization of $L(s)$ as $L(s)/\epsilon$ while preserving the form of Eq.~\eqref{eq:dW/dt full text}. This renormalization ensures that $L(s)\sim \mathcal{O}(1)$, consistent with the mean-field renormalization of $W_{\alpha\beta}$. For simplicity, we maintain the notation $L(s)$ for the renormalized function.

\subsection{Circuit Motif Derivation of the Evolution of Synaptic Weights } \label{sec:motif_derivation}

To derive a closed form dynamical system for ${\bf W_m}$ we only consider up to the second order motifs in Eq. \eqref{eq:dW/dt full text}. 

\subsubsection{Zeroth order motif}

The zeroth order motif is defined by $q^0_{EE}=p_0r_E^2$ and $M^0=1$. Thus, the coefficient $S^0$ is determined only by the integration of the STDP curve: $S^0=\int L(s)ds$. This term captures the synaptic plasticity owing to chance and unstructured pre- and post-synaptic spiking activity. Since the higher order terms involve synaptic connections then they scale with $\epsilon=1/N << 1$. Without any constraints on $L(s)$ the zeroth order term $p_0 r^2 S^0 \sim \mathcal{O}(1)$ (since $p_0 r^2 \sim \mathcal{O}(1)$) and would dominate the weight dynamics in Eq.~\eqref{eq:dW/dt full text} so that $W_{AB} \approx p_0 r^2 \epsilon^{-1}S^0 t $ \cite{ocker2015self}. This would lead to uninteresting synaptic dynamics, where only rapid assembly fusion or assembly decay are possible solutions. As a consequence, throughout our work we enforce the following constraint on $L(s)$:
\begin{equation} \label{eq:q_0}
    p_0 r^2 S^0 \sim \mathcal{O}(\epsilon) \Longrightarrow S^0=\int_{-\infty}^{\infty}L(s)ds \sim \mathcal{O}(\epsilon). 
\end{equation} 
This permits higher order terms in Eq.~\eqref{eq:dW/dt full text} to contribute to the growth or decay of the synaptic weights. 
Biologically, constraint \eqref{eq:q_0} amounts to the depression and potentiation components of the STDP rule canceling to $\mathcal{O}(\epsilon)$. 

\subsubsection{First order motifs}

The first order motifs include direct feedforward connections $\beta\rightarrow\alpha$ and feedback paths $\alpha\rightarrow \beta$ with corresponding response kernels: 
\begin{equation} \label{eq:q_1}
\begin{split}
    & q^f_{\alpha\beta}=\epsilon W_{\alpha\beta}; \hspace{0.3cm}M^f(s)  = K_{E}*C^0_{E}(s), \\
    &q^b_{\alpha\beta}=\epsilon p_0W_{\beta\alpha}; \hspace{0.3cm} M^b(s) = C^0_{E} * K^{-}_{E}(s). \\
\end{split}
\end{equation}
The negative superscript indicates reversed time $K^{-}(s)=K(-s)$ coming from the inverse Fourier transform of the complex conjugate. The $p_0$ in $q^b_{\alpha\beta}$ captures the probability of a backwards connection from populations $\beta$ to $\alpha$.   
The forwards and backwards first order coefficients $S^f_E$ and $S^b_E$ are then given by:
\begin{equation} \label{eq:S_1}
    \begin{split}
        S^f &= \int_{-\infty}^{\infty}L(s)\big( K_E * C_E^0 \big)(s)ds,\\
        S^b & =  \int_{-\infty}^{\infty}L(s)(C_E^0 *  K_E^-)(s)ds.\\
    \end{split}
\end{equation}


\subsubsection{Second order motifs}

The second order motifs include common inputs $\alpha \leftarrow \gamma \rightarrow \beta$ from another population $\gamma$, and feedforwad chains $\beta \rightarrow \gamma \rightarrow \alpha$ and backward chains $\alpha \rightarrow \gamma \rightarrow \beta$ through an intermediary population $\gamma$. The corresponding response kernels are respectively:
\begin{equation*}
\begin{split}
    M^{cc}_{\gamma}(s) &= (K_{E} * C^0_{\gamma} * K^-_{E})(s), \\
    M^{fc}_{\gamma}(s) &= (K_{E} * K_{\gamma} * C^0_{E})(s), \\
    M^{bc}_{\gamma}(s) &= (C^0_{E} * K_{\gamma}^{-} * K^{-}_{E}(s). \\
\end{split}
\end{equation*}
We distinguish the kernels by the intermediate population $\gamma \in \{E,I\}$. This is because while the homeostatic inhibitory plasticity makes it that excitatory populations $\alpha,\beta \in \{A,B,O\}$ all have the same response properties, those of the inhibitory and excitatory populations may differ (i.e in general we have $K_E(s) \ne K_I(s)$ and $C_E^0(s) \ne C_I^0(s)$). From these kernels, we compute the second order motif coefficients as:
\begin{equation} \label{eq:S_2}
    \begin{split}
        S^{cc}_\gamma &= \int_{-\infty}^{\infty}L(s)\big( K_E * C^0_\gamma * K^{-}_E \big)(s)ds,\\
        S^{fc}_\gamma & =  \int_{-\infty}^{\infty}L(s)(K_E * K_\gamma * C^0_E)(s)ds,\\
        S^{bc}_\gamma & =  \int_{-\infty}^{\infty}L(s)( C^0_E * K^{-}_\gamma * K^{-}_E)(s)ds.
    \end{split}
\end{equation}
In what follows, the coefficients for forward chain and backward chains always appear as a sum $S^{fb}=S^{fc}+S^{bc}$, and we use $S^{fb}$ for simplicity of exposition. 

Finally, the full second order motifs terms are:
\begin{equation} \label{eq:q_2}
\begin{split}
   &\sum\limits_{cc,fb} q^{\textrm{motif}}_{\alpha\beta}(\mathbf{W_m})=\\
    &+ \epsilon p_0(S^{cc} + S^{fb})\left[ \left (\frac{\gamma_E - \lambda}{2} \right) F_{\alpha\beta}({\bf W_m}) + \lambda W_{\alpha O}W_{O\alpha}\right]\\
        & + \epsilon  p_0 \gamma_I(S^{cc}_{I} W_{\alpha I}W_{\beta I} + S^{fb}_{I}W_{\alpha I}W_{IE}).\\
\end{split} 
\end{equation}
 The second order terms are distinguished by the function $F_{\alpha\beta}({\bf W})$ which captures the two pathway common inputs to the post-synaptic population $\alpha$ that originate in the pre-synaptic population $\beta$: 
\begin{align*}
F_{AA}({\bf W_m}) &= W_{AA}^2+W_{AB}^2; \nonumber \\
&\quad A \rightarrow A \rightarrow A \textrm{ and } A \rightarrow B \rightarrow A, \nonumber \\
F_{AB}({\bf W_m}) &= 2W_{AA}W_{AB}; \nonumber \\
&\quad B \rightarrow A \rightarrow A \hspace{0.2cm}(\times 2), \nonumber \\
F_{AO}({\bf W_m}) &= W_{AA}W_{AO}+W_{AB}W_{AO}; \nonumber \\
&\quad O \rightarrow A \rightarrow A \textrm{ and } O \rightarrow B \rightarrow A, \nonumber \\
F_{OA}({\bf W_m}) &= W_{AA}W_{OA}+W_{AB}W_{OA}; \nonumber \\
&\quad A \rightarrow A \rightarrow O \textrm{ and } A \rightarrow B \rightarrow O, \nonumber \\
F_{OO}({\bf W_m}) &= 2W_{OA}^2; \nonumber \\
&\quad O \rightarrow A \rightarrow O \hspace{0.2cm}(\times 2).
\end{align*}
In the above we have assumed symmetry in the weight evolution with $W_{BA}=W_{AB}$, $W_{OB}=W_{OA}$ and $W_{BO}=W_{AO}$. We remark that the term $\lambda W_{\alpha O} W_{O \alpha}$ results from the  $\alpha \rightarrow O \rightarrow \beta$ and $\beta \rightarrow O \rightarrow \alpha$ pathways and uses the above mentioned symmetries. 

The final term in Eq.~\eqref{eq:q_2} captures how correlated activity induced by common inhibition drives $W_{\alpha\beta}$ dynamics. In principle, we should have another two dynamical weights $W_{AI}$ and $W_{OI}$ and analyze a full seven dimensional mean field. However, as in \cite{ocker2019training}, we take the timescale of homeostatic inhibition to be much faster than that of excitatory learning ($1/\epsilon$). This allows inhibition to adiabatically track the excitatory weights and maintain control of the excitatory firing rate $r_E$. We can write $W_{AI}=W_{BI}$ by symmetry and $W_{OI}$ in terms of the excitatory weights and the steady state firing rates $r_E$ and $r_I$:    
\begin{equation} 
\begin{split}
    W_{AI} &= \mu^{\textrm{diff}}_A - \frac{r_E}{r_I \gamma_I}\Big[\big(\frac{\gamma_E - \lambda}{2}\big) (W_{AA}+W_{AB}) + \lambda W_{AO} \Big],\\
    W_{OI} &= \mu^{\textrm{diff}}_O - \frac{r_E}{r_I \gamma_I}\Big[(\gamma_E - \lambda) W_{OA} + \lambda W_{OO} \Big].\\
\end{split} \label{eq:Inh}
\end{equation}
Here, $\mu^{\textrm{diff}}_{\alpha}$ is a constant bias that determines the inhibitory weight strength needed for neuron population $\alpha$ to maintain its target firing rate $r^{\textrm{diff}}_{\alpha}$, as specified by the inhibitory plasticity rule, when receiving only external background input $r^{\textrm{ext}}_{\alpha}$ without excitatory input. A detailed derivation of inhibitory connection strength $W_{\alpha I}$ is provided in Methods sec.~\ref{appendix:inhibitory_plasticity}.

\subsubsection{Self consistent mean field theory for ${\bf W_m}$ }
Truncating Eq.~\eqref{eq:dW/dt full text} at second order and inserting Eqs.~\eqref{eq:q_0}, \eqref{eq:q_1} and \eqref{eq:q_2} gives the evolution equation for the generic weight $W_{\alpha\beta}$ ($\alpha,\beta \in \{A,B,O\}$) as:
\begin{equation}
\begin{split}
     & \frac{d W_{\alpha\beta}}{dt} 
         =  p_0 r_E^2 S^0/\epsilon + ( S^f  W_{\alpha\beta} + p_0 S^b \cdot W_{\beta\alpha})  \\  
         & +  p_0(S^{cc} + S^{fb})\left[ \left (\frac{\gamma_E - \lambda}{2} \right) F_{\alpha\beta}({\bf W_m}) + \lambda W_{\alpha O}W_{O\alpha}\right]\\
        & +   p_0 \gamma_I(S^{cc}_{I} W_{\alpha I}W_{\beta I} + S^{fb}_{I}W_{\alpha I}W_{IE}).\\
\end{split} \label{eq:GenMF}
\end{equation}
Recall that $S^0 \sim \mathcal{O}(\epsilon)$ (Eq.~\ref{eq:q_0}) so that, in principle, all terms on the right hand side of Eq.~\eqref{eq:GenMF} can contribute to the dynamics of ${\bf W_m}$. 

Eqs.~\eqref{eq:Inh} and \eqref{eq:GenMF} constitute a self consistent mean field theory for the evolution of mean synaptic weights $\bf{W_m}$. The influence of overlap parameter $\lambda$ is explicit in the second order terms in Eq.~\eqref{eq:GenMF}. However, the influence of the causality of the learning rule $L(s)$ given by $\kappa$ is more subtle. The derivation of Eq.~\eqref{eq:GenMF} is based on the separation of the fast timescale of spiking activity from the slow timescale of synaptic plasticity.  
As in many fast-slow systems analyses, the fast system is connected to the slow system via a temporal averaging of the fast system. This averaging is expressed in the calculations of the motif coefficients $S$ in Eqs.~\eqref{eq:q_0}, \eqref{eq:S_1} and \eqref{eq:S_2} where $L(s)$ (and hence $\kappa$) appear.

We remark that our mean field theory ignores that the operating point of the network about which we linearize in principle depends on the weight matrix ${\bf W_m}$, and thus the kernel $K_E(s)$ and the autocorrelation function $C^0_E(s)$ would also depend on $L(s)$. However, in practice the inhibitory plasticity ensures that the $E$ neurons firing rates remain at their target value irrespective of ${\bf W}$ (Appendix.~\ref{appendix:inhibitory_plasticity}, See also \cite{vogels2011inhibitory, ocker2019training}). Because of this homeostasis the linearization point is roughly independent of the recurrent weights, and thus $K_E(s)$ and $C^0_E(s)$ are also independent of $L(s)$ \cite{ocker2019training}. This simplifies the analysis of assembly fate considerably.

In the next section we use our mean field theory to study the $(\kappa,\lambda)$ dependence of assembly dynamics post training.

\begin{figure*}
    \begin{center}
    \vspace{-1mm}
    \setlength{\belowdisplayskip}{3pt}
    \setlength\abovedisplayskip{0pt}
    \includegraphics[scale=1]{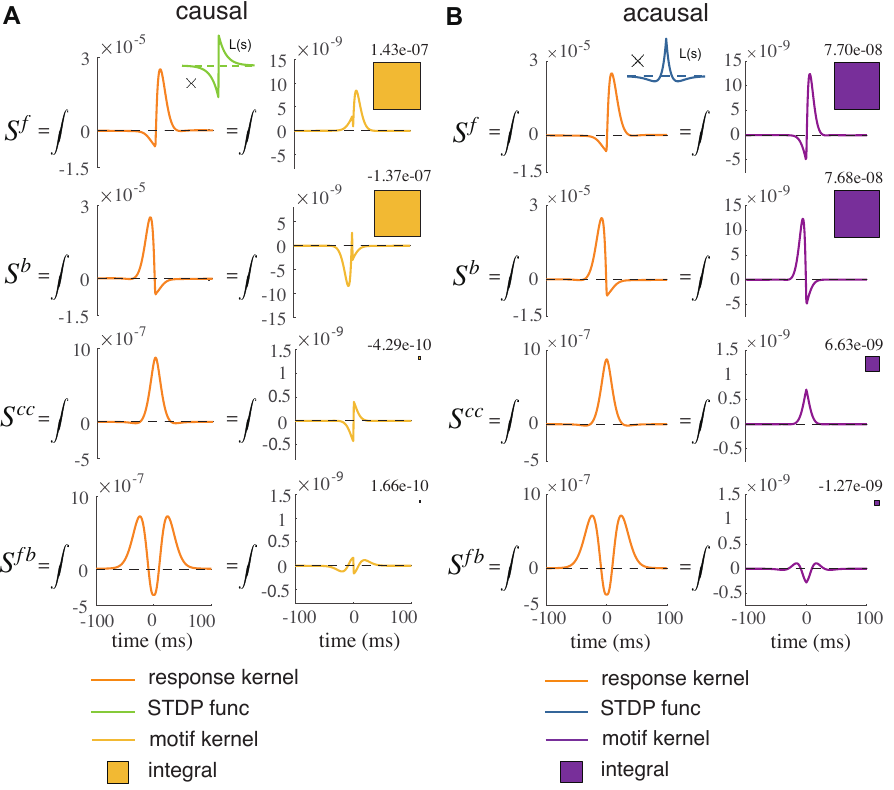}
    \end{center}
    \caption{Motif coefficients for assembly weight dynamics for the causal $(\kappa=1)$ and acausal $(\kappa=0)$ STDP rules. (A) Calculation of the $S$ coefficients for causal STDP. Each coefficient is the integral of the product of the STDP rule $L(s)=L^c(s)$ with the appropriate spike train covariance component (left plot). The yellow boxes are a relative comparison of the magnitude of the four coefficients. (B) same as (A) but for acausal STDP $L(s)=L^{ac}(s)$.  We note for the acausl rule that the second order motif coefficients ($S^{cc}$ and $S^{fb}$) are comparable to the first order coefficients ($S^f$ and $S^b$). The coefficients $S^{\rm{motif}}$ here are very small because we compute them in the fast timescale, where the unit is milliseconds. }\label{fig:ch3_fig5}
\end{figure*}

\subsection{How Motif Coefficients Depend on the Learning Rule}  \label{sec:Smotifs}

We begin our analysis by considering the forward motif coefficient $S^f$. The response/synaptic kernel $K(s)$ is weighted towards positive lag $s$, where the pre-synaptic neuron $j$ fires before the post-synaptic neuron $i$ (Fig.~\ref{fig:ch3_fig5}A, top).  The causal ($\kappa=1$) STDP rule $L(s)$ strictly potentiates for $s>0$, so that the combination of $K(s)$ and $L(s)$ leads to a large potentiation component (large yellow box in Fig.~\ref{fig:ch3_fig5}A, top). The backwards coefficient $S^b$ is the same as $S^f$ except that it is biased towards the spike train correlations where post-synaptic neuron $i$ fires before pre-synaptic neuron $j$, given by the kernel $K^-(s)$. This occurs for $s<0$ and results in a large depression component for the causal $L(s)$ rule (Fig.~\ref{fig:ch3_fig5}A, middle-top).

A similar argument applies for $S^f$ and $S^b$ with the acasual ($\kappa=0$) learning rule $L(s)$ (Fig.~\ref{fig:ch3_fig5}B, top and top-middle). The one distinction is that both $S^f$ and $S^b$ show potentiation (i.e $>0$) for the acausal rule. This is due to the temporal overlap of spike train correlations and the narrow potentation component of the acausal $L(s)$ for small $\vert s \vert$. However, the key distinction between assembly dynamics with causal or acausal learning occurs with the second order terms of our theory.  

Unlike the correlations through mono-synaptic forward and backward connections, the correlations in the spike trains from neurons $i$ and $j$ that are induced by common inputs are symmetric across lag $s$. This is true for the direct common input term $S^{cc}$ (Fig.~\ref{fig:ch3_fig5}A, B, middle-bottom) and when we combine the forward and backward chain terms $S^{fb}=S^{fc}+S^{bc}$, the effective correlations are also symmetric (Fig.~\ref{fig:ch3_fig5}A, B, bottom).  

The causal STDP rule ($\kappa=1$) is such that $L(s)$ is almost an odd function (in $s$). In this case, if $S^0=\int_{-\infty}^{\infty}L(s;\kappa=1)ds \sim \mathcal{O}(\epsilon)$, then when $L(s)$ is integrated against the even motif kernel functions $K^{cc}(s)=K_E * C^0_E * K^{-}_E(s)$ we obtain: 
\begin{equation*}
     S^{cc}=\int_{-\infty}^{\infty}K^{cc} (s) L(s;\kappa=1)ds \sim \mathcal{O}(\epsilon).    
\end{equation*}
This assumes that the timescales of $K^{cc}(s)$ are not much faster than those of $L(s)$ (i.e. $\tau_s \sim \tau_+ \sim \tau_-$) so that $\mathcal{O}(1)$ differences in $L(s)$ for $\pm s \to 0$ do not dominate $S^{cc}$.   The same is true for the chain term $S^{fb}$. 
Consequently, the coefficients $S^{cc}$ and $S^{fb}$ are much smaller than $S^f$ and $S^b$ (very small yellow boxes in Fig.~\ref{fig:ch3_fig5}A, middle-bottom and bottom). 
These small $S^{cc}$ coefficients imply that the dynamics of $W_{\alpha\beta}$ in Eq.~\eqref{eq:GenMF} is effectively linear, since the quadratic terms are negligible for reasonable values of $W_{\alpha\beta}$. 

By contrast, when the STDP learning rule is fully acausal ($\kappa=0$) then $L(s)$ is an even function (over $s$). Consequently, when $L(s)$ is integrated against the even and narrow  spike train correlations from common inputs we have: 
\begin{equation*}
    S^{cc}=\int_{-\infty}^{\infty}K^{cc} (s) L(s;\kappa=0)ds \sim \mathcal{O}(1) \hspace{0.5cm}   
\end{equation*}
This requires that $K^{cc}(s)$ primarily overlaps the potentiation component of $L(s)$; if that is the case then $S^{cc} \sim \mathcal{O}(1)$ even when $S^0  \sim \mathcal{O}(\epsilon)$. The result is sizable $S^{cc}$ and $S^{fb}$ coefficients (purple boxes in Fig.~\ref{fig:ch3_fig5}B, middle-bottom and bottom), especially for $S^{cc}$.  The same argument and conclusion holds for the coefficients $S^{cc}_I$ and $S^{fb}_I$. Thus, for acausal learning the quadratic terms in Eq.~\eqref{eq:GenMF} may contribute to the overall dynamics of assembly structure. We explore this possibility in the next section.
  
\begin{figure*}
    \begin{center}
    \vspace{-4mm}
    \setlength{\belowdisplayskip}{3pt}
    \setlength\abovedisplayskip{0pt}
    \includegraphics[scale=1]{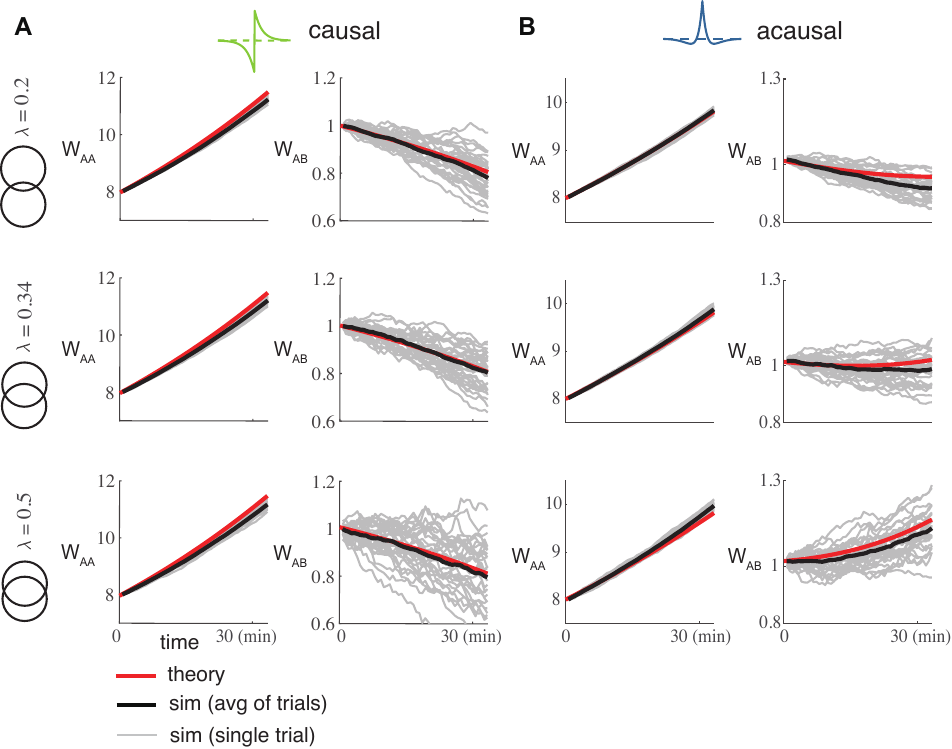}
    \end{center}
    \caption{Almost linear assembly weight dynamics for the causal STDP rule $(\kappa=1)$ and the nonlinear (quadratic) dynamics for acausal $(\kappa=0)$ STDP rule. (A) Comparison between theory and simulation for the within assembly weight $W_{AA}$ and between assembly weight $W_{AB}$ post training for causal STDP rule $L(s)=L^c(s)$. The grey curves are multiple realizations from the network of spiking neuron models; the black curve is the the average of these realizations; and the red curve is the theoretical mean field prediction. We perform this for three overlap $\lambda$ values (top to bottom). (B) Same as (A), but for acausal STDP rule $L(s)=L^{ac}(s)$. }  \label{fig:ch3_fig6}
\end{figure*}

\subsection{Causal/Acausal Learning Allows Assembly Segregation/Fusion for Large Overlap}

We have established that for the causal STDP rule the quadratic terms in Eq.~\eqref{eq:GenMF} are negligible (due to small coefficients $S^{cc}$ and $S^{fb}$). Since the overlap parameter $\lambda$ only appears in the quadratic terms, then for networks with causal STDP assembly overlap is expected to have minimal impact on post training assembly dynamics. In particular, if during training we embed segregated assemblies where $W_{AB} < W_{AA}$ and they persist post training for $\lambda=0$, then they should persist across a broad range of $\lambda>0$ as well.  This prediction is validated through simulations of the full spiking model network, as well as the associated mean field theory in Eqs.~\eqref{eq:Inh} and \eqref{eq:GenMF} and  across a range of $\lambda$ (Fig.~\ref{fig:ch3_fig6}A). We note the excellent agreement between the trial averaged simulations of our spiking network with plastic synapses ($N=1000$ and $p_0=0.1$ so approximately $1 \times 10^5$ synapses) and our five dimensional mean field theory in Eq.~\eqref{eq:GenMF} (compare red and black curves in Fig.~\ref{fig:ch3_fig6}A).   

By contrast to networks with causal learning, networks with acausal STDP have non-negligible coefficients for the quadratic (disynaptic) terms in the mean field theory in Eq.~\eqref{eq:GenMF}. 
For small $\lambda$ and if sufficient training is given so that $W_{AA}$ is significantly larger that $W_{AB}$ at the conclusion of training, then the assemblies will remain segregated post training (Fig.~\ref{fig:ch3_fig6}B, top). However, for larger overlap $\lambda$ the common term $\lambda W_{AO}W_{OA}$ in Eq.~\eqref{eq:GenMF} can dominate the disynaptic interactions, and this term is symmetric across the evolution of both $W_{AA}$ and $W_{AB}$. Consequently, this will promote growth for both within and between assembly weights, so that assembly fusion occurs (Fig.~\ref{fig:ch3_fig6}B, bottom).

In sum, we see that the influence of common inputs can cause assembly fusion if the learning rule $L(s)$ is acausal and the assembly overlap is large. We explore this result in the next section.  

\begin{figure*}
    \begin{center}
    \includegraphics[scale=1]{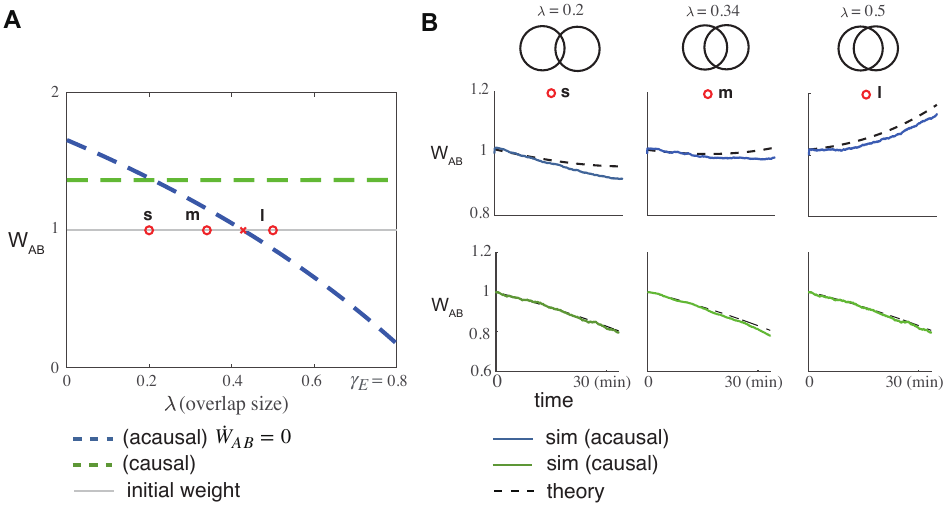}
    \end{center}
    \vspace{-4mm}
    \caption{ One dimensional dynamical system for the evolution of the between assembly weighy $W_{AB}$. (A) The threshold weight $W_{AB}^t$ that seperates growth and decay of $W_{AB}(t)$ as the overlap ratio $\lambda$ varies. The green dashed line is for the causal rule $L(s;\kappa=1)$ where the threshold is independent of $\lambda$. The blue dashed line is for the acausal rule $L(s;\kappa=0$) where as overlap $\lambda$ increases $W_{AB}^t$ is lowered. The grey line is the initial condition $W_{AB}(0)$ immediately after training. The three red dots with label s (small), m (medium), l (large) are the three overlaps considered in Fig. \ref{fig:ch3_fig5}. (B) The after training dynamics for $W_{AB}$ in our reduced system with the three overlaps considered in panel a). The dashed curves are from our reduced theory in Eq. \eqref{eq:1dMF} and the solid lines are from simulations of our full spiking network. The top row is for the acausal rule for $L(s)$, while the bottom row is for the causal rule.} \label{fig:ch3_fig7}
\end{figure*}

\subsection{The Boundary Between Assembly Segregation and Fusion} \label{sec:ch3_seg_fuse}

The fate of assembly training (segregation versus fusion) is determined from the growth or decay of the between assembly mean weight $W_{AB}$. To give deeper insight into the mechanics of assembly segregation/fusion we consider the approximated dynamical system where 
${\bf W_m}=[W_{AA},W_{AB}, W_{AO}, W_{OA}, W_{OO}]= [W_{\rm max},W_{AB}, W_{\rm max},W_{\rm max},W_{\rm max}]$. This amounts to assuming that the within and overlap assembly weights evolve to a large (maximal) value $W_{\rm max}$ and we are left considering only the one dimensional dynamics of the mean between assembly weights $W_{AB}$. Under this reduction we can write the following dynamics for $W_{AB}$: 
\begin{equation}
    \frac{dW_{AB}}{dt}=k_0+k_1W_{AB}+k_2W^2_{AB}. \label{eq:1dMF}
\end{equation}
Here the coefficients $k_0$, $k_1$ and $k_2$ are given by:
\begin{widetext}

\begin{equation*}
\begin{split}
k_0
=
p_0 r^2 S^0/ \epsilon + p_0 \gamma_I S^{cc}_I \bigl(D_0^2 \bigr) + p_0 \gamma_I S^{fb}_I \bigl( D_0 W_{IE} \bigr) + p_0 \bigl(S^{cc} + S^{fb}\bigr) \lambda W_{\rm max}^2.
\end{split}
\end{equation*}

\begin{equation*}
\begin{split}
k_1 = \bigl(S^f + p_0\,S^b\bigr) +
p_0 \gamma_I S^{cc}_I \bigl(-2\,D_0\,D_1\bigr)
+
p_0\gamma_I S^{fb}_I \bigl(-D_1\,W_{IE}\bigr)
\;+\;
p_0 \bigl(S^{cc} + S^{fb}\bigr) \bigl(\gamma_E - \lambda\bigr) W_{\rm max}.
\end{split}
\end{equation*}

\begin{equation*}
k_2 = p_0 \gamma_I S^{cc}_I D_1^2,
\end{equation*}

\end{widetext}
Here we introduce two hyperparameters $D_0, D_1$ for simplicity:
\begin{equation*}
\begin{split}
    D_0 &= \mu^{\mathrm{diff}}_{A} - \frac{r_E}{r_I\gamma_I}\Bigl[\frac{(\gamma_E - \lambda)(1+\lambda)}{2}\Bigr]W_{\rm max},\\
    D_1 &= \frac{r_E^{\rm{target}}(\gamma_E - \lambda)}{2 r_I \gamma_I}, 
\end{split}
\end{equation*}
where $D_1$ is a positive constant and $D_0$ is an intermediate variable that depends on $W_{\rm max}$. Both coefficients exhibit clear dependence on the overlap ratio $\lambda$. The $\mathcal{O}(1)$ scaling of these coefficients, combined with $W_{AB}\sim \mathcal{O}(1)$, ensures that all terms in Eq.~\eqref{eq:1dMF} maintain $\mathcal{O}(1)$ scaling. We assume that when training terminates (at $t=0$), the initial condition for the between-assembly weight is set to $W_{AB}(0) < W^{\rm max}$.

When the STDP learning $L(s)$ is causal ($\kappa=1$) then all second order motif can be ignored (i.e $S^{cc}, S^{fb}, S^{cc}_I$, and $S^{fb}_I$ are all approximately 0). This makes $k_2 \approx 0$, $k_1=S^f+p_0 S^b >0$ and $k_0=p_0 S^0 (r_E^{\rm{target}})^2 / \epsilon < 0$ (since $S^0<0$ for our choice of $L(s)$). For this system there is a unstable (threshold) point $W_{AB}^t=-k_0/k_1$, where for initial condition $W_{AB}(0)>W_{AB}^t$ then $W_{AB}(t)$ grows, while for $W_{AB}(0)<W_{AB}^t$ then $W_{AB}(t)$ decays. Importantly, for $\kappa=1$ we have that $k_0$ and $k_1$ are independent of the overlap parameter $\lambda$, and consequently so is the threshold $W_{AB}^t$ (Fig.~\ref{fig:ch3_fig7}A, green curve). Thus, for training that drives $W_{AB}(0)<W_{AB}^t$, then after training we will have assembly segregation independent of $\lambda$ (Fig.~\ref{fig:ch3_fig7}B; green curves).   



For acausal STDP learning ($L(s)$ with $\kappa=0$), the coefficients $k_0$ and $k_1$ take more complex forms and $k_2>0$. Specifically, all coefficients $k_0$, $k_1$ and $k_2$ depend on the overlap parameter $\lambda$ and second order motif coefficients $S^{\textrm{motif}}$ and by extension also on the causality parameter $\kappa$. Notably, the zeroth order motif coefficient $S^0<0$ remains largely independent of other terms in $k_0$, allowing us to select parameter regimes where $k_0<0$. Under these conditions, the quadratic equation $k_2 W_{AB}^2 + k_1 W_{AB} + k_0 = 0$ yields two real roots: one positive and one negative. Since $k_2 > 0$, the positive root is an unstable fixed point $W_{AB}^t$, again establishing a threshold between potentiation and depression of $W_{AB}$. 
In our model, $W_{AB}^t$ decreases with increasing $\lambda$, as the overlap population facilitates the growth of $W_{AB}$ (Fig.\ref{fig:ch3_fig7}A, blue curve). Consequently, at large overlap values $\lambda$, when $W_{AB}(0)>W_{AB}^t$, assembly fusion occurs (Fig.\ref{fig:ch3_fig7}B, blue curves). This relationship between overlap and fusion is robust across a broad range of parameters and is not specific to our parameter choice.

\begin{figure}[t!]
    \begin{center}
    \includegraphics[scale=1]{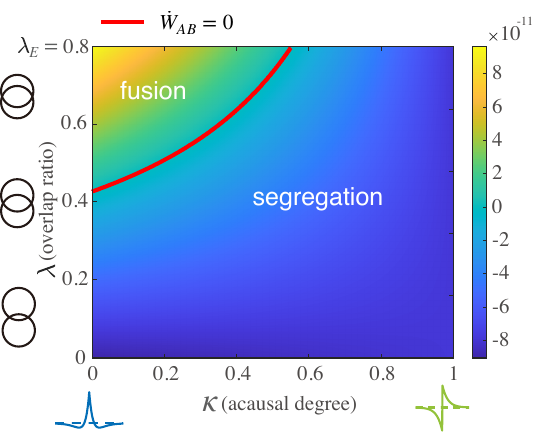}
    \end{center}
    \vspace{-4mm}
    \caption{Heat map of the value of the derivative $dW_{AB}/dt$ corresponding to different overlap parameter $\lambda$ and different causality parameter $\kappa$ with the initial weight $W_{AB}(0)$. The red curve specifies $d W_{AB}/dt$=0} \label{fig:ch3_fig8}
\end{figure}

Generally, for $\kappa$ not equal to $0$ or $1$, the STDP rule $L(s)=\kappa L^c(s) + (1-\kappa)L^{ac}(s)$ would be a combination of these two cases.  We notice that all the $S$ coefficients are linear in $L(s)$, and hence also linear in $\kappa$. Consequently, the mean weights dynamics are merely linear combinations of the dynamics for the weights when $\kappa=1$ and $\kappa=0$. Thus, for $\kappa \in (0,1)$ we do not expect new dynamics, and expect only the threshold $W_{AB}^t$ to be of interest.  For arbitrary $(\kappa,\lambda)$ we can compute the threshold $W_{AB}^t(\kappa,\lambda)$ and compare it to the initial condition: $W_{AB}(0)-W_{AB}^t(\kappa,\lambda)$. This gives us a boundary in $(\kappa,\lambda)$ parameter space that separates the assembly fates of fusion and segregation (Fig.~\ref{fig:ch3_fig8}, red curve). We see that for large overlap $\lambda$ then for sufficiently acausal $L(s)$ rules we have the fusion state. By contrast, for sufficiently causal $L(s)$ rules we will never have a fusion state even for $\lambda \to \gamma_E$ (of course for $\lambda=\gamma_E=0.8$ the notion of segregation versus fusion is moot).     

\section{Discussion}

Following previous work \cite{ocker2015self,ocker2019training,ravid2016shaping}, we derived a low dimensional dynamical system describing the mean weight dynamics of a weakly coupled network of spiking neuron models whose synapses obey a spike timing dependent plasticity (STDP) rule. We used this low-dimensional mean field theory to analyze how the stability of trained assembly structure depends on the combination of assembly overlap and the degree of causality in the STDP learning rule. A causal STDP rule suppresses fusion dynamics and allows assemblies to stay segregated regardless of the overlap size. This was not the case for acausal STDP rules, where for sufficient assembly overlap a fused assembly would develop after training. 

The key difference between assembly stability with causal versus acausal STDP rules is how they treat common input projections to neuron pairs between assemblies. The coefficients for the weight dynamics characterizing how the common input motifs affect the dynamics is computed as an integral of the product of the STDP rule and filtered spike train auto-covariance, of which the latter is an even function.  For causal STDP rules, which are close to odd functions, this product is an odd function so the integral is near zero. This makes the weight dynamics unaffected by the common inputs coming from the overlap group. By contrast, an acausal STDP curve is close to an even function, so that its product with auto-covariance is also even, and hence common input effects cannot be ignored. As a result fused assemblies can occur for sufficiently large assembly overlap.

\subsection{Rate-Based versus Timing-Based Synaptic Learning}

While STDP learning is inherently temporal, the degree of synaptic potentiation compared to depression is well known to depend on the firing rates of the pre- and post-synaptic neurons \cite{sjostrom2001rate,pfister2006triplets,graupner2016natural}. Biologically realistic STDP learning rules based on voltage or calcium dynamics capture the rate dependence of STDP \cite{clopath2010connectivity,graupner2012calcium}. When these realistic learning rules are used then the strong reciprocal coupling between neuron pairs that is the basis of assembly structure can be embedded in networks of spiking neuron models \cite{litwin2014formation,zenke2015diverse}. However, the complexity of these plasticity rules make a comprehensive theory of assembly formation in spiking systems difficult to formulate. Alternatively, models that consider simple, phenomological STDP rules which depend only on the relative timing of pre- and post-synaptic spiking are amenable to mean field treatments \cite{ocker2015self,gilson2010emergence,ravid2016shaping}. 


The mechanisms behind the formation and stability of assembly structure with rate-based \cite{litwin2014formation,zenke2015diverse} and timing-based plasticity are distinct. In a rate-based scenario, coordinated high firing rates for neurons within an assembly is a necessary condition for potentiation of within assembly synapses. However, due to balanced recurrent inhibition only a subset of assemblies can have access to high firing rates at a given time. A consequence of this is that  training can only be achieved through giving signals sequentially, so that
coordinated high firing rates can drive potentiation for synapses within the same assembly and low firing rates can drive depression for synapses between neuron pairs in different assemblies. In our temporal framework, the formation of the assembly structure depends on the strong spike time covariance within assemblies due to correlated stimuli (or feedforward inputs). Because of this the training of different assemblies can occur simultaneously, as opposed to sequentially. 

The stability of the assembly structure during spontaneous activity after training also differs between rate-based and timing-based learning. In the rate-based scenario, the trained structure is an multi-attractor state, and the assembly structure maintains itself through random activation of different assemblies (attractor states), ultimately reinforcing the learned structure \cite{litwin2014formation}. In the spike-timing-based scenario, it is the strong spike train correlations within assemblies due to strong recurrent connections that reinforce the learned structure. Hence, the assembly structure reinforces itself even when all neurons fire at approximately the same firing rate. 

A common criticism purely timing-based learning is that for non-trivial weight dynamics to occur we require the STDP curve to be roughly balanced between potentiation and depression (i.e $S^0=\int_{-\infty}^{\infty}L(s)ds \sim \mathcal{O}(\epsilon)$). This may occur for a restricted range of pre- and post-synaptic neuron firing rates \cite{sjostrom2001rate,litwin2014formation}, yet in general is not guaranteed. Rate-based or the recently coined behavioral timescale plasticity mechanisms \cite{magee2020synaptic} are more robust, and do not require fine tuning of the plasticity rule. Nonetheless, in all cases network learning requires access to both potentiation and depression mechanics, else all synaptic connections will become homogeneous in strength. The requirement that $S^0 \sim \mathcal{O}(\epsilon)$ for STDP learning simply has this fact being satisfied for statistically stationary spike trains. 

Finally, the STDP models employed in our work are phenomenological ones where $\Delta W$ is additive (Eq.~\ref{eq:delta W}), meaning that it does not depend on $W$. In general for such models weight dynamics either increases to the upper bound or depresses to zero \cite{rubin2001equilibrium,gutig2003learning}, which is inconsistent with unimodal and long-tailed weight distributions typically observed in experiments \cite{song2005highly, lefort2009excitatory}. Multiplicative STDP learning does not have this failure \cite{gutig2003learning}, however such a rule does not permit the motif-based mean field theory we use in our study \cite{ocker2015self}. Further modifications could be done by introducing axonal or dendritic delays or weight-dependence of plasticity which can yield weight distributions more closely resembling those observed in neural tissue \cite{babadi2010intrinsic, gutig2003learning, rubin2001equilibrium, rubin2001steady, gilson2011stability}. Additionally, our conclusion is made for the phenomenological shape of the STDP rule, the comparison to more realistic synaptic plasticity rules, such as triple STDP or calcium-based models \cite{pfister2006triplets, graupner2012calcium} remains to be analyzed.

\subsection{Overlap Capacity}

The stability of overlapping assemblies has been recently explored by Manz and colleagues \cite{manz2023purely} in networks with acausal excitatory STDP. Their work suggests that stable assembly structures with overlap can emerge under specific conditions: either when neurons maintain appropriate spontaneous firing rates or when most neurons participate in multiple assemblies with similar saturation levels. Our study is complementary and focuses on how the temporal structure of the STDP learning rule affects the maximal tolerated overlap. In particular, we have shown that overlap is much more tolerated when the STDP learning rule is causal, as opposed to the acausal rule used in \cite{manz2023purely}. 

In our study we only considered two symmetric assemblies with a common overlap, which is the simplest case. Applying the same derivation to the case of multiple assemblies with different population sizes would result in a larger dimensional mean field theory. We would need to allow for the case that $W_{AB} \ne W_{BA}$, and for a three population network we would have four overlap populations to model: $O_{AB}$, $O_{BC}$, $O_{AC}$ and $O_{ABC}$. However, all the dynamical equations for $W_{\alpha\beta}$ would have a similar form as our two assemblies case yet with extra terms corresponding to common input through weak connections from other assemblies. Nevertheless, the motif coefficients $S^{cc}, S^{fc}$, and $S^{bc}$ would still be the same. Hence our theoretical framework could be extended to multiple assemblies and our general conclusion likely still holds.

Considering multiple assemblies then begs the question of assembly capacity: how many assemblies of size $M<N$ can be embedded into a network of $N$ neurons and remain segregated post training? This question is distinct, but related, to more classic studies of memory capacity in recurrent networks \cite{aljadeff2021synapse,fusi2021memory}. For instance, in Hopfield networks capacity is measured by many input patterns can be stored in the recurrent weight matrix and reliably retrieved \cite{amit1985spin,krotov2023new}. In standard Hopfield networks all neurons contribute to the storage of each input pattern and hence the pattern `overlap' is significant (or complete). In our network model capacity could be measured by an assembly population's ability to perform pattern completion and pattern segregation of inputs that are coherent with the training set \cite{knierim2016tracking,litwin2014formation}. Assembly fusion would certainly compromise pattern segregation, as now inputs to assembly $\alpha$ would also activate assembly $\beta$ through the strong $\alpha \to \beta$ wiring. This research avenue will be pursued in future studies.       

\section{Methods}

\subsection{Diffusion-based theory for first and second order spike train statistics} \label{diffusion_methods}

We present the method used to derive spike train firing rates and cross-correlations in our network model. For large networks, the recurrent input to neuron $i$ in population $\alpha$ can be treated using a diffusion approximation:
\begin{equation*}
\begin{split}
    I_\alpha^i(t) &= \sum_{\beta=E,I}\sum_{j=1}^{N^\beta}W_{\alpha\beta}^{ij}(t)\cdot \big(J_{\textrm{syn}} * y^j_\beta(t)\big) \\
    & \approx \mu_{\rm{rec}}^i(t) + \sigma_{\rm{rec}}^i(t) \xi^i(t).
\end{split}
\end{equation*}
Here we drop the population $\alpha$ notation for ease of exposition. In brief, the recurrent input is replaced by a white noise process with time-varying noise strength (time-changed Brownian motion). The external input $I^i_{\rm{ext}}$ is similarly modeled as white noise (Wiener process) $I^i_{\rm{ext}} = \mu_{\alpha}^{\rm{ext}} + \sigma_{\alpha}^{\rm{ext}} \xi^i_{\alpha}(t)$ (see Eq.~\ref{eq:ch2_Iext}). The sum of these two independent stochastic processes yields a white noise with time-varying strength \cite{oksendal2013stochastic}, resulting in the voltage equation for neuron $i$ in population $\alpha$ (modified from Eq.~\ref{eq:ch2_EIF}):
\begin{equation} \label{eq:ch2_FP_volequ}
\begin{split}
    \tau \frac{dV^i}{dt} =& (E_L - V^i) + \Delta_T \exp\Big(\frac{V^i - V_T}{\Delta_T}\Big) + \mu^i(t)\\
    &+ \sigma^i(t)\sqrt{2\tau} \xi^i(t).
\end{split}
\end{equation}

The mean input $\mu(t)$ and noise strength $\sigma(t)$ are given by:
\begin{equation*}
\begin{split}
    \mu^i(t) &= \mu_{\rm{ext}} + \mu^i_{\rm{rec}}(t), \\
    \sigma^i(t) &=  \sqrt{\frac{\sigma_{\rm{ext}}^2 +  (\sigma^i_{\rm{rec}})^2(t)}{2\tau}} \approx \frac{\sigma_{\rm{ext}}}{\sqrt{2\tau}}.
\end{split}
\end{equation*}
The approximation for $\sigma^i(t)$ holds because $\sigma^i_{\rm{rec}}\sim \mathcal{O}(\epsilon)$ when recurrent connections are weak.

The voltage equation is a Langevin equation, whose probability distribution $P(V,t)$ obeys a corresponding Fokker-Planck equation:
\begin{equation*}
    \frac{\partial P}{\partial t}(V^i,t)=\frac{\partial P}{\partial V^i}\left (V^i-\mu(t)+\frac{\sigma^2(t)}{2}\frac{\partial P}{\partial V^i} \right)
\end{equation*}
with boundary conditions:
\begin{equation*}
\begin{split}
P(V_{\text{th}-}, t) &= P(V_{\text{re}+}, t) - P(V_{\text{re}-}, t) = 0, \\
J(V_{\text{th}-}, t) &= J(V_{\text{re}+}, t) - J(V_{\text{re}-}, t) = r^i(t).
\end{split}
\end{equation*}
These conditions arise from the threshold and reset mechanism, with $r^i(t)$ being the firing rate of neuron $i$ at time $t$.

To compute temporal mean firing rates and cross-correlations, we assume the network operates near a steady state with small fluctuations. This approach separates the input to each neuron into static and dynamic components:
\begin{equation*}
    \mu^i(t)=\overline{\mu}^i + \widetilde{\mu}^i(t),  
\end{equation*}
where $\overline{\mu}^i$ represents the time-averaged input and $\widetilde{\mu}^i(t)$ satisfies $\int_0^\infty\widetilde{\mu}^i(s)ds=0$. We again assume the variance remains approximately static, $\sigma^i \approx \sigma_{\rm{ext}}$.

Following \cite{richardson2007firing, richardson2008spike} we could compute the steady-state firing rates $r^i_0$ given the mean input $\overline{\mu}^i$ and diffusion parameter $\sigma_{\rm{ext}}$:
\begin{equation*}
    r^i_{0} = f(\overline{\mu}^i, \sigma_{\rm{ext}})
\end{equation*}
by numerically solving the Fokker-Planck equation. Since neurons within each subpopulation $\alpha$ share identical parameters except for $\mu^i$, their transfer functions are equivalent:
\begin{equation*}
    r^i_{0} = f_{\alpha}(\overline{\mu}^i, \sigma_{\textrm{ext}})
\end{equation*}
for neuron $i$ in subpopulation $\alpha\in\{A,B,O,I\}$.

The mean input current $\overline{\mu}^i$ depends on both external and recurrent inputs, which in turn depend on the recurrent weights $\mathbf{W}$ and firing rates $\mathbf{r}_0$. This leads to a self-consistent system of equations:
\begin{equation*}
    \mathbf{r}_{0} = \mathbf{f}(\mathbf{r}_{0}; \mathbf{W},\sigma_{\rm{ext}})
\end{equation*}
where the weights $\mathbf{W}$ evolve slowly enough to be treated as static parameters.

The firing rates can be obtained by integrating the auxiliary differential equations \cite{layer2022nnmt}:
\begin{equation*}
    \dot r^i = - r^i + f_{\alpha}(\mu^i, \sigma_{\textrm{ext}})
\end{equation*}
until reaching a fixed point $\dot r_i = 0$, starting from an initial guess $r_{\textrm{initial}}$. This method effectively estimates individual neuron firing rates under weak connections embedded in white noise. Our theoretical predictions align with simulation results, particularly in showing that homeostatic PV plasticity drives excitatory neurons toward the target rate $r_E^{\rm{target}}$.

To compute the spike train cross-covariances we employ a linear response analysis of the Fokker-Planck equation. When presented with a small input perturbation $I_{\epsilon}(t)$, the firing rate response can be approximated as $I_{\epsilon} * \chi_r(t)$, where $\chi_r(t)$ is the response kernel computed numerically from the Fokker-Planck equation \cite{richardson2007firing, richardson2008spike}.

In the frequency domain this becomes $I_{\epsilon}(\omega) \chi_r(\omega)$, for frequency variable $\omega$. For small fluctuations $\tilde{\mu}^i(t)$, we can apply this to mean-corrected spike trains \cite{trousdale2012impact, ocker2019training}:
\begin{equation*}
    y^i(\omega)= \int_{-\infty}^\infty (y^i(t) - r^i_{0}) e^{-2 \pi i \omega t}dt.
\end{equation*}
Following \cite{lindner2005theory, trousdale2012impact} we decompose the spike train as:
\begin{equation}\label{eq:3}
\begin{split}
    y^i(\omega) &= y^i_{0}(\omega) + \chi^i(\omega)\widetilde{\mu}^i(\omega) \\
                &= y^i_{0}(\omega) + \chi^i(\omega)\sum_{j=1}^{N}W^{ij} J_{\textrm{syn}}(\omega)y^j(\omega),
\end{split}
\end{equation}
where $\chi^i(\omega)$ represents the Fourier transform of the response kernel and $J_{\textrm{syn}}(\omega)$ is the Fourier transform of the synaptic filter.

For matrix notation simplicity, we define the effective interaction matrix \cite{trousdale2012impact, ocker2019training}:
\begin{equation} \label{eq:ch2_defK}
    \mathbf{K}(\omega) = \Big[ \chi^i(\omega) J^{ij}_{\textrm{syn}}(\omega) \Big]
\end{equation}
allowing us to rewrite Eq.~$\eqref{eq:3}$ as:
\begin{equation*}
    \begin{split}
        \mathbf{y} & = \mathbf{y}_{0} + \mathbf{W}\circ \mathbf{K}(\omega)\cdot\mathbf{y}\\
        \mathbf{y} & = \big(\mathbf{I} - \mathbf{W}\circ \mathbf{K}(\omega)\big)^{-1}\cdot\mathbf{y}_{0}
    \end{split}
\end{equation*}

The cross-covariance in the frequency domain is then:
\begin{equation} \label{eq:ch2_CrossCov}
\begin{split}
    \mathbf{C}(\omega) & = \big\langle\mathbf{y}(\omega)\cdot\mathbf{y}^{*}(\omega)\big\rangle \\
    & = \big(\mathbf{I} - \mathbf{W}\circ \mathbf{K}(\omega)\big)^{-1} \mathbf{C}_{0}(\omega)
    \big(\mathbf{I} - (\mathbf{W}\circ \mathbf{K}(\omega))^*\big)^{-1}
\end{split}
\end{equation}
where $\mathbf{C}_0(\omega)$ is the Fourier transform of the mean-subtracted auto-correlation. Given that weights $\mathbf{W}$ evolve on a much slower timescale than spiking activity, we can treat them as static when computing $\mathbf{C}(s;\mathbf{W})$. Our simulation results confirm that this linear response approach provides accurate estimates of cross-covariance in our network model.

\subsection{Mean field dynamics of population averaged synaptic weights}  \label{methods:equations}

In this section we derive the mean field equations of all the averaged weights ${\bf W_m}=[W_{AA}, W_{AB}, W_{AO}, W_{OA}, W_{OO}].$ After training the network is dissected into four populations $\{A,B,O,I\}$. We recall that we have introduced the overlap ratio $\lambda = \frac{N_O}{N}$, and excitatory and inhibitory ratios $\gamma_E = \frac{N_E}{N}, \gamma_I = \frac{N_I}{N}$. Notice here population $O$ contains all the neurons in the overlap between two assemblies, and population $A$ or $B$ contain only the neurons in the assembly $A$ or $B$ respectively, but not in the overlap part. Hence the number of neurons in these two populations equals to $N_{A} = N_{B} = \frac{\gamma_E - \lambda}{2}N$.

We define the mean variables for synaptic weights from population $\beta$ to $\alpha$ as:
\begin{equation*}
    W_{\alpha\beta} = \frac{1}{N_\alpha N_\beta}\sum_{i\in \alpha}\sum_{j\in \beta}\frac{W^{ij}_{\alpha\beta}}{\epsilon} \sim \mathcal{O}(1).
\end{equation*}
We apply the methods introduced in Sec.~\ref{sec:derivation} to expand the cross-covariance in its Neumann Series and truncate at the second order motifs. We take the term corresponding to the common chain $i\leftarrow k \rightarrow j$ motifs (cc) as an example, which evaluates to:

\begin{equation*}
    \begin{split}
        \int_{-\infty}^{\infty}&L(s)C^{\textrm{cc}}(s)ds = \\
        &\epsilon \cdot p_0 S^{cc} \Big[ \big(\frac{\gamma_E - \lambda}{2}\big) (W_{\alpha A}W_{\beta A} + W_{\alpha B}W_{\beta B}) + \\ &\lambda W_{\beta O}W_{\beta O} + \gamma_I W_{\alpha I}W_{\beta I} \Big]. 
    \end{split}
\end{equation*}
Other terms are also approximated by ordinary differential equations of $\bf{W_m}$ with parameter $\lambda$.

Applying the same process and we have a set of mean-field equations of $\bf{W_m}$:
\begin{widetext}
\begin{equation*}
    \begin{split}
        \frac{d W_{AB}}{dt} &= p_0 r^2 S^0 / \epsilon + W_{AB} (S^f + p_0 S^b) + p_0 \gamma_I(S^{cc}_I W_{AI}^2 + S^{fb}_I W_{AI}W_{IE})\\
                & + p_0(S^{cc}+S^{fb})\Big[ (\gamma_E - \lambda) W_{AA}W_{AB} + \lambda W_{AO}W_{OA}\Big]
    \end{split}
\end{equation*}

\begin{equation*}
    \begin{split}
        \frac{d W_{AA}}{dt} &= p_0 r^2 S^0 / \epsilon + W_{AA} (S^f + p_0 S^b) + p_0 \gamma_I(S^{cc}_I W_{I}^2 + S^{fb}_I W_{AI}W_{IE}) \\
                & + p_0(S^{cc}+S^{fb})\Big[ \Big(\frac{\gamma_E - \lambda}{2}\Big) (W_{AA}^2+W_{AB}^2) + \lambda W_{AO}W_{OA}\Big] 
    \end{split}
\end{equation*}

\begin{equation*}
    \begin{split}
    \frac{d W_{OO}}{dt} &= p_0 r^2 S^0 / \epsilon + W_{OO} (S^f + p_0 S^b) +  p_0 \gamma_I(S^{cc}_I W_{OI}^2 + S^{fb}_I W_{OI}W_{IE})\\
    & + p_0(S^{cc}+S^{fb})\Big[(\gamma_E - \lambda) W_{OA}^2 + \lambda W_{OO}^2\Big]\\
    \end{split}
\end{equation*}

\begin{equation*}
    \begin{split}
    \frac{d W_{AO}}{dt} &= p_0 r^2 S^0 / \epsilon + (S^fW_{AO} S^f + p_0 S^b W_{OA}) +  p_0 \gamma_I(S^{cc}_I W_{AI}W_{OI} + S^{fb}_I W_{AI}W_{IE})\\
    & + p_0(S^{cc}+S^{fb})\Big[\Big(\frac{\gamma_E - \lambda}{2}\Big) (W_{AA}W_{AO}+W_{AB}W_{AO}) + \lambda W_{AO}W_{OO}\Big]\\
    \end{split}
\end{equation*}

\begin{equation*}
    \begin{split}
    \frac{d W_{OA}}{dt} &= p_0 r^2 S^0 / \epsilon +  (S^f W_{OA} + p_0 S^b W_{AO}) +  p_0 \gamma_I(S^{cc}_I W_{AI}W_{OI} + S^{fb}_I W_{OI}W_{IE})\\
    & + p_0(S^{cc}+S^{fb})\Big[\Big(\frac{\gamma_E - \lambda}{2}\Big) (W_{AA}W_{OA}+W_{AB}W_{OA}) + \lambda W_{AO}W_{OA}\Big]\\
    \end{split}
\end{equation*}
\end{widetext}
And the inhibitory synaptic weights follow the linear relationship as described in Methods~\ref{appendix:inhibitory_plasticity}:
\begin{widetext}
\begin{equation*}
\begin{split}
    W_{AI} &= \frac{\mu_E^{\textrm{target}} - \mu_A^{\rm{ext}} }{r_I \gamma_I} - \frac{r_E}{r^p \gamma_I}\Big[\big(\frac{\gamma_E - \lambda}{2}\big) (W_{AA}+W_{AB}) + \lambda W_{AO} \Big],\\
    W_{OI} &= \frac{\mu_E^{\textrm{target}} - \mu_O^{\rm{ext}} }{r_I \gamma_I}  - \frac{r_E}{r^p \gamma_I}\Big[(\gamma_E - \lambda) (W_{OA}) + \lambda W_{OO} \Big].\\
\end{split}
\end{equation*}
\end{widetext}
Here, the first term $(\mu_E^{\textrm{target}} - \mu_A^{\rm{ext}})/(r_I \gamma_I)$, denoted as $\mu^{\textrm{diff}}_{E}$ for brevity before, determines the inhibitory weight strength required for excitatory neurons to maintain their target firing rate $r^{\textrm{diff}}_{\alpha}$ when receiving only external background input $r^{\textrm{ext}}_{\alpha}$ without excitatory input. This relationship is maintained through homeostatic inhibitory plasticity \cite{vogels2011inhibitory, ocker2019training}. A detailed derivation is provided in Methods~\ref{appendix:inhibitory_plasticity}.





\subsection{Inhibitory STDP and homeostatic control of firing rates} \label{appendix:inhibitory_plasticity}

In this section, we describe how inhibitory synaptic weights from $I$ neurons to $E$ neurons ($W_{EI}^{ij}$) evolve under STDP. The inhibitory plasticity rule combines two components: a timing-dependent function and pre-synaptic spike-induced depression.

The homeostatic STDP function is defined for lag $s=t_{\rm{post}}-t_{\rm{pre}}$ as:
\begin{equation*}
    L^h(s) = f^h \exp\Big({-\frac{|s|}{\tau^h}}\Big).
\end{equation*}
Here $f^h<0$ is the update step size and $\tau^h$ is the STDP decay time constant. Additionally, each pre-synaptic spike triggers synaptic depression with magnitude $d^h>0$:
\begin{equation*}
    W^{ij}_{EI}\rightarrow W^{ij}_{EI} + d^h.
\end{equation*}
This plasticity model induces a target firing rate $r_E^{\textrm{target}}$, defined by the relationship between depression and potentiation:
\begin{equation*}
    r_E^{\textrm{target}} = -\frac{d^h}{2 f^h \tau^h}.
\end{equation*}
$r_E^{\textrm{target}}>0$ is ensured by $d^h>0$ and $f^h<0$. After averaging over neuronal populations, the weight evolution equation reduces to:
\begin{equation*}
    \frac{dW_{\alpha I}}{dt} = p_{EI}S^0_{EI} \cdot r_I \big[r_E^{\textrm{target}} - r_{\alpha}\big] + \sum_{\textrm{motif}} S^{\textrm{motif}} \cdot q^\textrm{motif}_{\alpha\beta}(\mathbf{W_m})
\end{equation*}
where $S^0_{EI} = 2f^h\tau^h < 0 \sim \mathcal{O}(1)$. Since the motif terms are $\mathcal{O}(\epsilon)$, the leading-order dynamics depend primarily on the firing rates $r_I$ and $r_E$.

The mean firing rate of excitatory neurons $\alpha \in \{A,B,O\}$ follows:
\begin{equation*}
\begin{aligned}
    r_{\alpha} &= f_E(\mu_{\alpha} , \sigma_{\alpha} )\\
    \mu_{\alpha} &= \mu_{\textrm{ext}} + \sum_{\gamma} N_{\gamma} W_{\alpha\gamma} r_{\gamma} + N_{I} W_{\alpha p} r_I
\end{aligned}
\end{equation*}
where $f_E(\mu, \sigma)$ is a monotonically increasing transfer function. For a given noise intensity $\sigma_{\textrm{ext}}$, there exists an input $\mu_E^{\textrm{target}}$ such that $r_E^{\textrm{target}} = f_E(\mu_E^{\textrm{target}}, \sigma_{\textrm{ext}})$.
Analysis of the dynamics shows that:
\begin{equation*}
    \frac{dW_{\alpha I}}{dt} = \Big( p_{\alpha I} S^0_{EI} r_h \cdot f'(\tilde{\mu},\sigma) \Big) \Big(\mu_{\alpha} - \mu_E^{\textrm{target}}\Big) + \mathcal{O}(\epsilon).
\end{equation*}
Given that $W_{\alpha I} \sim \mathcal{O}(\epsilon)$, this equation implies:
\begin{equation*}
    r_{\alpha} - r_{E}^{\textrm{target}} \sim \mathcal{O}(\epsilon).
\end{equation*}
This convergence occurs through the balanced interaction between inhibitory input $N_I W_{\alpha I} r_I$ and excitatory input $\sum_{\gamma} N_{\gamma} W_{\alpha\gamma} r_{\gamma}$.

The $I$ neuron firing rate $r_I$ remains static due to non-plastic connections $W_{I\alpha}$ and $W_{II}$:
\begin{equation*}
\begin{aligned}
    \mu_{I} &= \mu_{\rm{ext}} + \sum_{\alpha\in E} N_{\alpha} W_{I \alpha} r_{E}^{\textrm{target}} + N_I W_{II} r_I\\
    r_I &= f_I (\mu_I, \sigma_{\textrm{ext}})
\end{aligned}
\end{equation*}
Consequently, the inhibitory connection weights can be directly computed:
\begin{equation*}
   W_{\alpha I} = \bigg(\mu_{e}^{\textrm{target}} - \mu_{\rm{ext}} - \sum_{\gamma\in E} N_{\gamma} W_{\alpha\gamma} r_{\gamma}\bigg)/(N_I r_{I}) 
\end{equation*}
This relationship demonstrates that inhibitory connections track excitatory weights linearly while maintaining target firing rates, a result we verified through numerical simulations.

\newpage

\subsection{Model parameters} \label{parameters}

\begin{table}[h]
\begin{ruledtabular}
\begin{tabular}{cccccc}
Param & Value & Unit & Param & Value & Unit\\
\hline
$\tau$ & 20 & ms & $N$ & 1000 & - \\
$V_L$ & -55 & mV & $p_0$ & 0.1 & -\\
$\Delta_T$ & 3 & mV  & $N_E$ & 800 & -  \\
$V_{\rm th}$ & -53 & mV & $N_I$ & 200 & - \\
$V_{\rm re}$ & -60 & mV & $W_{EE}^{\rm max}$ & 0.24 & mV \\
$\tau_{\rm ref}$ & 1 & ms & $W_{EI}^{\rm max}$ & -0.96 & mV\\
$\tau_{\rm syn}$ & 5 & ms & $W_{IE}$ & 0.08 & mV \\
$\mu$ & 0 & mV & $W_{II}$ & -0.32 & mV \\
$\sigma$ & 4 & mV \\
$\sigma_{\rm train}$ & 1.5 & mV \\
\end{tabular}
\end{ruledtabular}
\caption{\label{tab:neuronal}
Neuronal Parameters
}
\end{table}

\begin{table}[h]
\begin{ruledtabular}
\begin{tabular}{ccc}
Parameter & Value & Unit \\
\hline
$S^c_0$ & $-W_{EE}^{\rm max}*8.33*10^{-5}$ & mV \\
$\tau^c_{-}$ & 20\footnotemark[1] & ms\\
$\tau^c_{+}$ & 20\footnotemark[1] & ms \\
$f^c_{-}$ & $-W_{EE}^{\rm max}*3.33*10^{-3}$ & mV \\
$f^c_{+}$ & solved for consistency & - \\
\hline
$S^{ac}_0$ & $W_{EE}^{\rm max}*8.33*10^{-3}$ & mV \\
$\tau^{ac}_{-}$ & 30 & ms \\
$\tau^{ac}_{+}$ & 30 & ms \\
$f^{ac}_{-}$ & $W_{EE}^{\rm max}*1.11*10^{-4}$\footnotemark[2] & mV\\
$f^{ac}_{+}$ & $W_{EE}^{\rm max}*1.11*10^{-4}$\footnotemark[2] & mV \\
$T^{ac}_{+}$ & 30 & ms \\
$T^{ac}_{-}$ & solved for consistency & - \\
\hline
$\tau^h$ & 30 & ms \\
$f^h$ & $-W_{EE}^{\rm max}*6*10^{-4}$ & mV \\
$r^{\rm target}_E$ & 12 & Hz \\
$d^h$ & solved for consistency & -  \\
\end{tabular}
\end{ruledtabular}
\footnotetext[1]{For our analytical framework, we set $\tau^c_{-}=\tau^c_{+}$ to maintain $L^c(s)$ as approximately odd. In Fig.~\ref{fig:2} and~\ref{fig:ch3_fig3}, we used faster potentiation ($\tau^c_{+}=15$ ms) than depression ($\tau^c_{-}=30$ ms) to accelerate training. The causality and anti-symmetry of the STDP function preserve the validity of our theory under these conditions. This STDP configuration can be approximated within our framework using $\kappa\approx0.75$ and $\tau^{ac}_{-}=\tau^{ac}_{+}=20$ ms, thus remaining consistent with our theoretical analysis.}
\footnotetext[2]{These values, in combination with $T^{ac}_{+/-}$, ensure comparable magnitudes between the causal and acausal STDP functions $L^{c}(s)$ and $L^{ac}(s)$.}
\caption{\label{tab:plasticity}
Plasticity Parameters
}
\end{table}

\section{Acknowledgments}

B.D. is supported by the NIH grants 1U19NS107613-01, R01NS133598 
and CRCNS-R01EY034723, a Vannevar Bush faculty fellowship (N000141812002), and the Simons Foundation Collaboration on the Global Brain. This research benefited from Physics Frontier Center for Living Systems funded by the National Science Foundation (PHY-2317138). Financial support was provided via the National Institute from Mathematics and Theory in Biology (Simons Foundation award MP-TMPS-00005320 and NSF award DMS-2235451).

\bibliography{main}

\end{document}